\documentclass[11pt]{article}
\usepackage[svgnames,dvipsnames]{xcolor}
\usepackage{physics}
\usepackage{here}
\usepackage{amsmath,amsthm,amssymb,mathtools,bm}
\usepackage{latexsym}
\usepackage{multicol}
\usepackage{subfig}
\usepackage{overpic}
\usepackage{extarrows}
\usepackage{tikz}
\usepackage{xcolor}
\definecolor{Myblue}{rgb}{0,0,0.9}

\usepackage[all]{xy}

\numberwithin{equation}{section}
\usepackage{enumitem}

\usepackage[title,titletoc]{appendix}
\usepackage[height=8.85in,width=6.45in]{geometry}

\usepackage{tcolorbox}
\tcbuselibrary{breakable, skins, theorems}

\theoremstyle{plain}

\newtheorem*{thm*}{Theorem}

\newtheorem*{claim*}{Claim}

\makeatletter
\renewcommand{\appendix}{\par
  \setcounter{section}{0}%
  \setcounter{subsection}{0}%
  \gdef\presectionname{\appendixname}%
  \gdef\postsectionname{}%
  \gdef\thesection{\presectionname\@Alph\c@section\postsectionname}%
  \gdef\thesubsection{\@Alph\c@section.\@arabic\c@subsection}%
  \renewcommand{\theequation}{\@Alph\c@section.\arabic{equation}}%
  \renewcommand{\thefigure}{\@Alph\c@section.\arabic{figure}}%
  \renewcommand{\thetable}{\@Alph\c@section.\arabic{table}}%
}
\makeatother

\newcommand{\rmA}{\mathrm{A}}
\newcommand{\rmB}{\mathrm{B}}
\newcommand{\rmC}{\mathrm{C}}
\newcommand{\calA}{\mathcal{A}}
\newcommand{\calN}{\mathcal{N}}

\newcommand{\Rep}{\mathrm{Rep}}
\usepackage{./jheppub}

\title{
 $\bm{N}$-ality symmetry and SPT phases in (1+1)d
}

\author[1]{Jun Maeda,}
\affiliation[1]{Department of Physics,
Kyoto University, Kyoto, 606-8502, Japan}
\emailAdd{maeda@gauge.scphys.kyoto-u.ac.jp}

\author[2]{Tsubasa Oishi}
\affiliation[2]{Yukawa Institute for Theoretical Physics,
Kyoto University, Kyoto, 606-8502, Japan}
\emailAdd{tsubasa.oishi@yukawa.kyoto-u.ac.jp}
\preprint{YITP-25-35, KUNS-3038}

\abstract{
Duality symmetries have been extensively investigated in various contexts, playing a crucial role in understanding quantum field theory and condensed matter theory. In this paper, we extend this framework by studying $N$-ality symmetries, which are a generalization of duality symmetries and are mathematically described by $\mathbb{Z}_N$-graded fusion categories. In particular, we focus on an $N$-ality symmetry obtained by gauging a non-anomalous subgroup of $\mathbb{Z}_N\times\mathbb{Z}_N\times\mathbb{Z}_N$ symmetry with a type III anomaly. We determine the corresponding fusion rules via two complementary approaches: a direct calculation and a representation-theoretic method. As an application, we study the symmetry-protected topological (SPT) phases associated with the $N$-ality symmetry. We classify all such SPT phases using the SymTFT framework and explicitly construct lattice Hamiltonians for some of them.
}
\begin{document}
 \frenchspacing 
\maketitle

\section{Introduction}

Symmetry is one of the fundamental tools for analyzing quantum many body systems. In recent years, the concept of symmetry has been generalized \cite{Gaiotto:2014kfa}, leading to significant progress in understanding symmetries that cannot be expressed as unitary operators on a Hilbert space. These symmetries, known as non-invertible symmetries, are described by fusion categories in two-dimensional cases \cite{Bhardwaj:2017xup,Chang:2018iay}. For recent reviews, see \cite{Shao:2023gho, Schafer-Nameki:2023jdn}. One of the most classical examples of non-invertible symmetries is the Kramers-Wannier duality at the critical point of the Ising model \cite{Kramers:1941kn}. This symmetry is an example of a type of non-invertible symmetries known as duality symmetries, which are characterized by a topological manipulation forming group $\mathbb{Z}_2$ and are described by Tambara-Yamagami categories \cite{TAMBARA1998692}. In the case of the Ising model, the spin-flip $\mathbb{Z}_2$ symmetry defect and the duality symmetry defect form the $\mathrm{TY}(\mathbb{Z}_2)$ symmetry.

Anomaly matching conditions, just as in the case of conventional symmetries, can be applied to non-invertible symmetries as well. In particular, if a symmetry in the UV theory is anomalous, a unique gapped ground state is prohibited in the IR theory. For instance, since the $\mathrm{TY}(\mathbb{Z}_2)$ symmetry is anomalous \cite{Chang:2018iay, Thorngren:2019iar}, the ground state must either be degenerate or gapless\footnote{In the case where the topology of the spatial manifold is trivial, there is another possibility where the IR theory becomes a non-trivial TQFT with a unique gapped ground state. This is possible in higher dimensions.}. Indeed, the Ising model is gapless at its critical point.

Gapped phases with non-invertible symmetries have been extensively studied in recent years (see, e.g., 
\cite{Bhardwaj:2023fca,Bhardwaj:2023idu,Bhardwaj:2024qiv,Bhardwaj:2025piv}).
The simplest type of gapped phases is Symmetry protected topological (SPT) phase \cite{Chen:2010zpc,Chen:2011pg,Chen:2010gda,Gu:2009dr,Levin:2012yb,Pollmann:2009mhk,Pollmann:2009ryx}. SPT phases are symmetric gapped phases with a unique ground state on any closed manifold. Mathematically, they are classified by fiber functors of the corresponding fusion category \cite{Inamura:2021wuo,Thorngren:2019iar}. For example, $\Rep(D_8)$ admits three fiber functors, corresponding to three distinct SPT phases and the lattice Hamiltonians of these SPT phases on a tensor product Hilbert space were constructed in \cite{Seifnashri:2024dsd}.

A generalization of duality symmetry is $N$-ality symmetry, which is characterized by topological manipulations forming group $\mathbb{Z}_N$ \cite{Lu:2022ver,Lu:2024lzf,Ando:2024hun,Lu:2024ytl,Lu:2025gpt}. Duality symmetry, which is described by the Tambara-Yamagami category corresponds to the special case where $N=2$. $N$-ality symmetries, described by the $\mathbb{Z}_N$-graded fusion categories~\cite{Etingof:2009yvg,Gelaki:2009blp}, are studied in various contexts \cite{Thorngren:2021yso,Choi:2022zal,Choi:2022jqy,Cordova:2022ieu, Hayashi:2022fkw}. If $N$ is prime, the fusion rules of $N$-ality symmetry are a straightforward generalization of duality symmetry. However if $N$ is composite, the fusion rules of $N$-ality symmetry become significantly intricate.

In this paper, we consider an $N$-ality symmetry which can be obtained by gauging a non-anomalous subgroup of $\mathbb{Z}_N\times\mathbb{Z}_N\times\mathbb{Z}_N$ symmetry with a type III anomaly. This is a straightforward generalization of $\Rep(D_8)$ symmetry. In section \refeq{section 2}, we first present the general construction of non-invertible symmetry defects from anomalous group-like symmetries. We then review the case of $\Rep(D_8)$ symmetry, which is the simplest case \cite{Kaidi:2023maf,Thorngren:2019iar}. Finally, we generalize these results to the case of the $N$-ality symmetry and compute the fusion rules in two different approaches. The first approach is the field theoretical one. In this approach, we directly compute the fusion rules of symmetry defects expressed as functionals of gauge fields. This approach is useful in that it can also be applied to other $N$-ality symmetries, however when $N$ is composite, especially when it has a perfect square as a divisor, it becomes difficult to compute the fusion rules in this approach. The second approach is the representation theoretical one. The $N$-ality symmetry we discuss can be expressed as the Rep-category of non-Abelian group $G$ and thus symmetry defects can be regarded as Wilson lines, or equivalently characters of irreducible representations of $G$. We explicitly derive the expression as characters of the representations from the formulation of the defects as functionals of the gauge fields and compute the fusion rules. This method can only be applied to $N$-ality symmetries that can be written as a Rep-category, however it allows us to compute the fusion rules for arbitrary $N$.

In section \refeq{section 3}, we study SPT phases for the $N$-ality symmetry. Using the SymTFT perspective, we obtain a one-to-one correspondence between gapped phases of the $N$-ality symmetry and those of $\mathbb{Z}_N\times\mathbb{Z}_N\times\mathbb{Z}_N$ symmetry with a type III anomaly. We derive the conditions for the gapped phase of the $\mathbb{Z}_N\times\mathbb{Z}_N\times\mathbb{Z}_N$ symmetry corresponding to the SPT phases for the $N$-ality symmetry. As a result, we find that the $N$-ality symmetry admits $\sum_k(\mathrm{gcd}(N,k))^2$ distinct SPT phases for odd $N$ and $\sum_k\frac{3}{4}(\mathrm{gcd}(N,k))^2$ distinct SPT phases for even $N$, where the sum runs over $k$ such that $\left(\mathrm{gcd}(N,k)\right)^2\equiv 0 \mod N$. We also construct the lattice Hamiltonians for some of these SPT phases. We show that these SPT phases are in the same phase as SPT phases for $\mathbb{Z}_N\times\mathbb{Z}_N$ symmetry, which is a sub-symmetry of the $N$-ality symmetry by explicitly constructing an interface Hamiltonian between two SPT phases.


\section{Non-invertible symmetries obtained from group-like symmetries} \label{section 2}
\subsection{$A$-graded fusion categories} \label{sec 2.1}
In this section, we consider an Abelian group-like global symmetry $A\times B$ in a $(1+1)$d theory. Suppose there is a mixed anomaly between $A$ and $B$, with the $(2+1)$d anomaly inflow action given by
\begin{equation}\label{eq:anomaly inflow action}
    \calA = \int a\cup e(b),
\end{equation}
where $e\in H^2(B;\widehat{A})$\footnote{$\widehat{A}=\mathrm{Hom}(A;U(1))$ is the Pontryagin dual of $A$, which is isomorphic to $A$ for finite Abelian group $A$.}, and $a, b$ are background gauge fields for $A$ and $B$, respectively. Here, $e(b)$ is an $\widehat{A}$-valued 2-cocycle on the spacetime $X$, constructed from $e$ and $b$\footnote{More precisely, $e(b)$ can be defined as follows. There is a group extension labeled by $e\in H^2(B;\widehat{A})$ and this extension induces a short exact sequence of cochain complexes. Then a map $H^1(X;B)\to H^2(X;\widehat{A})$ is induced by the zigzag lemma and $e(b)$ is the image of $b$ of this map.}.

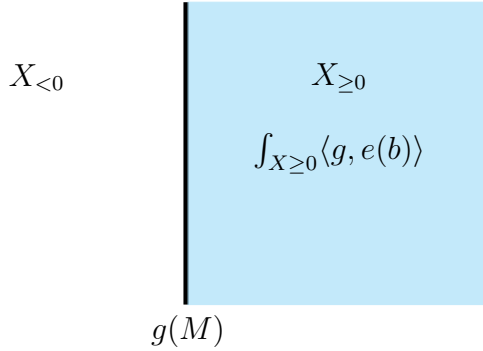
\begin{figure}[H]
    \centering 
    \begin{tikzpicture}
    \coordinate (A) at (1,0);
    \coordinate (XL) at (-1,3);
    \coordinate (XR) at (3,3);

   \draw[white] (-3,0)--(-3,4);
   \draw[line width = 2][black] (0.975,0)--(0.975,4);
    
    \filldraw[cyan!30,opacity=0.7] (1,0) rectangle (5,4);
    
    \node at (3,2) {\large$\int_{X\ge0} \langle g, e(b) \rangle$};
    \node at (A) [below] {\large $g(M)$};
    \node at (XL) {\large $X_{<0}$};
    \node at (XR) {\large $X_{\ge0}$};
\end{tikzpicture} 
    \caption{Anomaly inflow from the right bulk SPT phase to $M$.}
    \label{figure 1}
\end{figure}

Although $A\times B$ is anomalous, it is possible to gauge $A$ or $B$ since these subgroups are individually non-anomalous.
If we gauge $A$, it is known that the symmetry of the gauged theory is not simply $\widehat{A}\times B$ but rather a group extension of $B$ by $\widehat{A}$, classified by $e$ \cite{Tachikawa:2017gyf}. When gauging $B$, however, the resulting symmetry is typically more intricate and, in general, becomes non-invertible \cite{Tachikawa:2017gyf,Nguyen:2021yld,Heidenreich:2021xpr}. This can be understood as follows. If we gauge $B$, a symmetry defect of $A$ becomes non-topological due to the mixed anomaly. The anomaly inflow action \eqref{eq:anomaly inflow action} implies that, when we deform a defect of $g\in A$ on a line $M$ to that on $M'$, we acquire a phase
\begin{equation}\label{eq:phase from mixed anomaly}
    \int_{V} \langle g,e(b) \rangle ,
\end{equation}
where $V$ is the surface enclosed by $M$ and $M'$, and $\langle\cdot,\cdot\rangle$ denotes the canonical paring between $A$ and $\widehat{A}$. Another interpretation of the anomaly \eqref{eq:anomaly inflow action} is that, when we place a defect of $g$ on $M$, we should add an SPT phase
\begin{equation}\label{eq:SPT phase on half-space}
    \int_{X\ge0} \langle g, e(b) \rangle
\end{equation}
on the half-space $X_{\ge0}$ adjacent to $M$ (see figure \ref{figure 1}). The phase \eqref{eq:phase from mixed anomaly} then arises from the change of the half-space $X_{\ge0}$, which is the integration region in \eqref{eq:SPT phase on half-space}. In this interpretation, the non-topological nature of $g(M)$ can be understood as the anomaly inflow of the SPT phase \eqref{eq:SPT phase on half-space}.
To restore the topologicalness of $g(M)$, we must introduce additional degrees of freedom localized on $M$ that cancel the anomaly from the SPT phase. However, introducing these new degrees of freedom generally renders the defect non-invertible. These non-invertible defects are labeled by nontrivial elements of group $A$. Together with defects of $\widehat{B}$, whose label corresponds to the trivial element of $A$, these defects form an $A$-graded fusion category\footnote{The $A$-graded fusion category $\mathcal{C}$ is a fusion category which admits a direct sum decomposition $\mathcal{C}=\bigoplus_{g\in A}\mathcal{C}_g$, where the tensor product is defined as $\otimes : \mathcal{C}_g\times \mathcal{C}_h\rightarrow\mathcal{C}_{gh}$, $\forall g,h\in A$. When a defect $\calN$ of $\mathcal{C}$ is an element of $\mathcal{C}_g$, we define the grading of $\calN$ to be $g$.}. 
Such constructions of non-invertible symmetries have been studied in various contexts (see e.g. \cite{Tachikawa:2017gyf,Kaidi:2023maf, Choi:2022jqy, Kaidi:2021xfk}).


\subsection{Duality symmetry}
In this subsection, we give one of the simplest example among the types of non-invertible symmetries discussed in the previous subsection. First we consider the symmetry
\begin{equation}
    \mathbb{Z}_2^{\rmA}\times\mathbb{Z}_2^{\rmB}\times\mathbb{Z}_2^{\rmC}
\end{equation}
with a type III anomaly\footnote{Here, $\mathbb{Z}_2^{\rmA}$ corresponds to $A$ and $\mathbb{Z}_2^{\rmB}\times\mathbb{Z}_2^{\rmC}$ corresponds to $B$ in the previous subsection.}, whose anomaly inflow action is given by
\begin{equation}
    \calA = \pi\int a\cup b\cup c,
\end{equation}
where $a,b,c$ are $\mathbb{Z}_2^{\rmA},\mathbb{Z}_2^{\rmB},\mathbb{Z}_2^{\rmC}$ background gauge fields, respectively.

If we gauge $\mathbb{Z}_2^{\rmB}\times\mathbb{Z}_2^{\rmC}$, it is known that the symmetry of the gauged theory is $\Rep(D_8)$ symmetry \cite{Thorngren:2019iar,Kaidi:2023maf}. We review this for the generalization to the $N$-ality symmetry in the next subsection. 
We denote the original theory and the gauged theory as $\mathcal{T}$ and $\widehat{\mathcal{T}}$, respectively. The partition function of the gauged theory is given by
\begin{equation}
    Z_{\widehat{\mathcal{T}}}[B,C]=\frac{1}{|H^1(X;\mathbb{Z}_2)|}\sum_{b,c\in H^1(X;\mathbb{Z}_2)}Z_{\mathcal{T}}[b,c]\ e^{i\pi\int b\cup C+c\cup B},
\end{equation}
where $B$ and $C$ are background gauge fields for the dual symmetries $\widehat{\mathbb{Z}}_2^{\rmB}$ and $\widehat{\mathbb{Z}}_2^{\rmC}$, respectively. As explained above, the gauged theory has $\Rep(D_8)$ symmetry. To see this, we define two topological manipulations $\mathbf{S}$ and $\mathbf{T}$ as
\begin{equation}
\begin{aligned}
    &Z_{\mathbf{S}\mathcal{T}}[B,C]=\frac{1}{|H^1(X;\mathbb{Z}_2)|}\sum_{b,c\in H^1(X;\mathbb{Z}_2)}Z_{\mathcal{T}}[b,c]\ e^{i\pi\int b\cup C+c\cup B},\\
    &Z_{\mathbf{T}\mathcal{T}}[B,C]=Z_{\mathcal{T}}[B,C]\ e^{i\pi\int B\cup C}.
\end{aligned}
\end{equation}
$\mathbf{S}$ and $\mathbf{T}$ represent the gauging of $\mathbb{Z}_2^{\rmB}\times\mathbb{Z}_2^{\rmC}$ symmetry and the stacking of $\mathbb{Z}_2^{\rmB}\times\mathbb{Z}_2^{\rmC}$ SPT, respectively. They satisfy the relations $\textbf{S}^2=1$ and $\textbf{T}^2=1$\footnote{The equality of topological manipulations corresponds to the equality of maps between two theories.}. The non-invertible symmetry defect of the gauged theory is generated by a topological manipulation $\mathbf{TST}$. One can check this as follows: 
\begin{align}
    Z_{g\mathbf{TST}\widehat{\mathcal{T}}}[B,C]&=\frac{1}{|H^1(X;\mathbb{Z}_2)|^2}\sum_{b,c,\widetilde{b},\widetilde{c}\in H^1(X;\mathbb{Z}_2)}Z_{\mathcal{T}}[b,c]\ e^{i\pi\int b\cup c+b\cup\widetilde{c}+c\cup\widetilde{b}+\widetilde{b}\cup \widetilde{c}+\widetilde{b}\cup C+\widetilde{c}\cup B+B\cup C}\\ \nonumber
    &=\frac{1}{|H^1(X;\mathbb{Z}_2)|}\sum_{b,c\in H^1(X;\mathbb{Z}_2)}Z_{\mathcal{T}}[b,c]\ e^{i\pi\int b\cup C+c\cup B}\\
   &=Z_{\widehat{\mathcal{T}}}[B,C] \nonumber,
\end{align}
where $g$ is a global $\mathbb{Z}_2^{\rmA}$ transformation and acts on the partition function of $\mathcal{T}$ as $Z_{\mathcal{T}}[b,c]\rightarrow Z_{\mathcal{T}}[b,c]\ e^{i\pi\int b\cup c}$ due to the mixed anomaly. Thus, a symmetry defect of $\widehat{\mathcal{T}}$ can be constructed from the topological manipulation $\mathbf{TST}$ and it is a duality defect since $(\textbf{TST})^2=1$ \cite{Kaidi:2023maf}.

Next, let us construct a duality defect implementing $\mathbf{TST}$. As explained in subsection \refeq{sec 2.1}, the $\mathbb{Z}_2^{\rmA}$ defect becomes a duality defect by adding the degree of freedom which cancels the anomaly. In this case, it is sufficient to couple the $\mathbb{Z}_2^{\rmA}$ defect with $1$d TQFT
\begin{equation}\label{eq:Z_2 1d TQFT}
    \frac{1}{|C^0(M;\mathbb{Z}_2)|}\sum_{\phi_1,\phi_2\in C^0(M;\mathbb{Z}_2)}\exp\left[\pi i\int \phi_1c - \phi_2b - \phi_1\delta\phi_2\right],
\end{equation}
where $\phi_1,\phi_2$ are $\mathbb{Z}_2$-valued 0-cochain supported on $M$\footnote{There are other possible choices to cancel the anomaly; however it is known that such TQFTs can be decomposed into the product of the unique (so-called minimal) TQFT associated with the anomaly and a decoupled TQFT. A similar argument for 3d TQFTs is discussed in \cite{Hsin:2018vcg}. Our choice of TQFT \eqref{eq:Z_2 1d TQFT} is the minimal one, since the fusion rules \eqref{eq:fusion rules of Rep(D_8)} discussed below agree with those \eqref{eq:fusion rules of Rep(G)} obtained from the mathematical argument. Similar arguments apply to the 1d TQFTs in \eqref{eq:1d TQFT for N_1} and \eqref{eq:Z_d 1d TQFT}.}. We can check this as follows. Under the gauge transformation
\begin{equation}
    \begin{cases}
        b \rightarrow b + \delta\beta \\
        c \rightarrow c + \delta\gamma \\
        \phi_1 \rightarrow \phi_1 + \beta \\
        \phi_2 \rightarrow \phi_2 + \gamma,
    \end{cases}
\end{equation}
where $\beta,\gamma\in C^0(M;\mathbb{Z}_2)$, the partition function of the 1d TQFT \eqref{eq:Z_2 1d TQFT} is multiplied by
\begin{equation}
    \exp\left[\pi i\int \beta c-\gamma b -\gamma\delta \beta\right].
\end{equation}
Therefore, the 1d TQFT \eqref{eq:Z_2 1d TQFT} is anomalous and its anomaly is canceled by the anomaly inflow action
\begin{equation}
    \pi\int b\cup c.
\end{equation}
This implies the anomaly which arises when we put the $\mathbb{Z}_2^{\rmA}$ defect can be canceled by the 1d TQFT \eqref{eq:Z_2 1d TQFT}. Thus, the duality defect is given by
\begin{equation}
    \calN_1(M)=\frac{1}{|C^0(M;\mathbb{Z}_2)|}\sum_{\phi_1,\phi_2\in C^0(M;\mathbb{Z}_2)}g(M)\exp\left[\pi i\int \phi_1c - \phi_2b - \phi_1\delta\phi_2\right],
\end{equation}
where $g(M)$ is the $\mathbb{Z}_2^{\rmA}$ defect. The fusion rules are 
\begin{equation}\label{eq:fusion rules of Rep(D_8)}
\begin{aligned}
\calN_1\times\eta_b&=\eta_b\times\calN_1=\calN_1\\
\calN_1\times\eta_c&=\eta_c\times\calN_1=\calN_1\\
\calN_1\times\calN_1&=1+\eta_b+\eta_c+\eta_b\eta_c,
\end{aligned}
\end{equation}
where $\eta_b$ and $\eta_c$ are Wilson lines of $b$ and $c$, given by
\begin{equation}
    \eta_b=\exp\left[i\pi\oint b\right],\quad\eta_c=\exp\left[i\pi \oint c\right],
\end{equation}
respectively. These fusion rules coincide with those of $\Rep(D_8)$\footnote{More precisely, we cannot identify this symmetry category as $\Rep(D_8)$ based solely on the fusion rules. To make this identification, we must determine the data of the bicharacter and the Frobenius-Schur indicator of the symmetry category. However, we can identify it as $\Rep(D_8)$ since the symmetry of the gauged theory is $D_8$ when we gauge $\mathbb{Z}_2^{\rmA}$ in $\mathcal{T}$.}. 

\subsection{$N$-ality symmetry} \label{sec 2.3}
In this subsection, we discuss the generalization to $N$-ality symmetries. More precisely, we first consider the symmetry
\begin{equation}
  \mathbb{Z}_N^{\rmA}\times\mathbb{Z}_N^{\rmB}\times\mathbb{Z}_N^{\rmC}
\end{equation}
with a type III anomaly, whose anomaly inflow action is given by
\begin{equation} \label{type III anomaly}
    \mathcal{A} = \frac{2\pi}{N}\int a\cup b\cup c.
\end{equation}
Gauging $\mathbb{Z}_N^{\rmB}\times\mathbb{Z}_N^{\rmC}$ symmetry, we obtain a $\mathbb{Z}_N^{\rmA}$-graded fusion category symmetry and we call it $N$-ality symmetry.
In the following, we construct $N$-ality defects, basically following the same discussion as in the case of $N=2$. We again denote the original theory and the gauged theory as $\mathcal{T}$ and $\widehat{\mathcal{T}}$, respectively. The partition function of $\widehat{\mathcal{T}}$ is given by
\begin{equation}
    Z_{\widehat{\mathcal{T}}}[B,C]=\frac{1}{|H^1(X;\mathbb{Z}_N)|}\sum_{b,c\in H^1(X;\mathbb{Z}_N)}Z_{\mathcal{T}}[b,c]\ e^{\frac{2\pi i}{N}\int b\cup C+c\cup B},
\end{equation}
where $B$ and $C$ are background gauge fields for dual symmetries $\widehat{\mathbb{Z}}_N^{\rmB}$ and $\widehat{\mathbb{Z}}_N^{\rmC}$, respectively.

Then we define two topological manipulations $\mathbf{S}$ and $\mathbf{T}$ as
\begin{equation}
\begin{aligned}
    &Z_{\mathbf{S}\mathcal{T}}[B,C]=\frac{1}{|H^1(X;\mathbb{Z}_N)|}\sum_{b,c\in H^1(X;\mathbb{Z}_N)}Z_{\mathcal{T}}[b,c]\ e^{\frac{2\pi i}{N}\int b\cup C+c\cup B} \\
    &Z_{\mathbf{T}\mathcal{T}}[B,C]=Z_{\mathcal{T}}[B,C]\ e^{-\frac{2\pi i}{N}\int B\cup C}.
\end{aligned}
\end{equation}
$\mathbf{S}$ and $\mathbf{T}$ represent the gauging of $\mathbb{Z}_N^{\rmB}\times\mathbb{Z}_N^{\rmC}$ symmetry and the stacking of $\mathbb{Z}_N^{\rmB}\times\mathbb{Z}_N^{\rmC}$ SPT, respectively. They satisfy the relations $\textbf{S}^2=1$ and $\textbf{T}^N=1$. The non-invertible symmetry of the gauged theory is generated by a topological manipulation $\mathbf{STS}$\footnote{In the case of $N=2$, the relation $\mathbf{STS=TST}$ holds.}. One can check explicitly that the partition function of $\widehat{\mathcal{T}}$ is invariant under $\mathbf{STS}$.
\begin{equation}
\begin{aligned}
    Z_{g\mathbf{STS}\widehat{\mathcal{T}}}[B,C]&=\frac{1}{|H^1(X;\mathbb{Z}_N)|^3}
    \sum_{\substack{b,c,b',c',\widetilde{b},\widetilde{c}\\\in H^1(X;\mathbb{Z}_N)}}Z_{\mathcal{T}}[b,c]\ e^{\frac{2\pi i}{N}\int b\cup c+b\cup c'+c\cup b'+b'\cup\widetilde{c}+c'\cup\widetilde{b}-\widetilde{b}\cup\widetilde{c}+\widetilde{b}\cup C+\widetilde{c}\cup B}\\
    &=\frac{1}{|H^1(X;\mathbb{Z}_N)|}\sum_{b,c\in H^1(X;\mathbb{Z}_N)}Z_{\mathcal{T}}[b,c]\ e^{\frac{2\pi i}{N}\int b\cup C+c\cup B}\\
    &=Z_{\widehat{\mathcal{T}}}[B,C],
\end{aligned}
\end{equation}
where $g$ is a generator of global $\mathbb{Z}_N^{\rmA}$ symmetry and acts on the partition function of $\mathcal{T}$ as $Z_{\mathcal{T}}[b,c]\rightarrow Z_{\mathcal{T}}[b,c]\ e^{\frac{2\pi i}{N}\int b\cup c}$ due to the mixed anomaly. Thus, a symmetry defect of $\widehat{\mathcal{T}}$ can be constructed from the topological manipulation $\mathbf{STS}$ and it is a $N$-ality defect since $(\mathbf{STS})^N=1$.

To construct the $N$-ality defect, we need to attach a 1d TQFT which cancels the anomaly from the bulk to the defect $g$. The appropriate 1d TQFT is given by
\begin{equation}\label{eq:1d TQFT for N_1}
    \frac{1}{|C^0(M;\mathbb{Z}_N)|}\sum_{\phi_1,\phi_2\in C^0(M;\mathbb{Z}_N)}\exp\left[\frac{2\pi i}{N}\int \phi_1c - \phi_2b - \phi_1\delta\phi_2\right].
\end{equation}
Therefore, the $N$-ality defect is given by
\begin{equation}
    \calN_1(M) = \frac{1}{|C^0(M;\mathbb{Z}_N)|}\sum_{\phi_1,\phi_2\in C^0(M;\mathbb{Z}_N)}g(M)\exp\left[\frac{2\pi i}{N}\int \phi_1c - \phi_2b - \phi_1\delta\phi_2\right],
\end{equation}
where $g(M)$ is the $\mathbb{Z}_N^{\rmA}$ defect. The quantum dimension of this defect is $N$ since this 1d TQFT has $N$ degrees of freedom.

Next, we consider the construction of non-invertible defects with higher gradings. Naively, we can obtain non-invertible defects with grading $k$ by considering the topological manipulation $(\mathbf{STS})^k$, however there is a better choice when $N$ and $k$ are not coprime. Namely, when $\gcd(N,k)>1$, it is sufficient to couple $g(M)^k$ with a 1d TQFT with fewer degrees of freedom. Such a 1d TQFT is given by
\begin{equation}\label{eq:Z_d 1d TQFT}
    \frac{1}{|C^0(M;\mathbb{Z}_d)|}\sum_{\phi_1,\phi_2\in C^0(M;\mathbb{Z}_d)}\exp\left[\frac{2\pi i}{d}\int r\phi_1c-\phi_2b-\phi_1\delta\phi_2\right],
\end{equation}
where $d=\frac{N}{\gcd(N,k)}$ and $r=\frac{dk}{N}$. Indeed, under the gauge transformation
\begin{equation}\label{eq:gauge trfm. for Z_d}
    \begin{cases}
        b \rightarrow b + \delta\beta \\
        c \rightarrow c + r^{-1}\delta\gamma \\
        \phi_1 \rightarrow \phi_1 + \beta \\
        \phi_2 \rightarrow \phi_2 + \gamma,
    \end{cases}
\end{equation}
where $\beta,\gamma\in C^0(M;\mathbb{Z}_d)$\footnote{Although $\beta,\gamma$ usually take values in $\mathbb{Z}_N$, in the presence of only the defect $g^k(M)$, the partition function is gauge-invariant if $\beta,\gamma$ are multiples of $d$.} and $r^{-1}$ is the multiplicative inverse of $r$ modulo $d$, the 1d TQFT \eqref{eq:Z_d 1d TQFT} is anomalous, whose anomaly inflow action is given by
\begin{equation}
    \frac{2\pi k}{N}\int b\cup c.
\end{equation}
Therefore, we can define the non-invertible defect with grading $k$ as
\begin{equation}\label{eq:general N_k}
    \calN_k(M) = \frac{1}{|C^0(M;\mathbb{Z}_d)|}\sum_{\phi_1,\phi_2\in C^0(M;\mathbb{Z}_d)}g^k(M)\exp\left[\frac{2\pi i}{d}\int r\phi_1c-\phi_2b-\phi_1\delta\phi_2\right],
\end{equation}
where $d=\frac{N}{\gcd(N,k)}$ and $r=\frac{dk}{N}$, and the quantum dimension of $\calN_k$ is $d$.

When $\gcd(N,k)>1$, we can construct the other non-invertible defects with grading $k$ by attaching Wilson lines of $b$ and $c$\footnote{Since gradings of Wilson lines of $b,c$ are zero, the grading of a non-invertible defect is unchanged by attaching them.}. Indeed,
\begin{equation}\label{eq:N_k,s,t}
    \begin{split}
        \calN_k(M)\eta_{b}^s(M)\eta_c^t(M) = &
        \frac{1}{|C^0(M;\mathbb{Z}_d)|}\sum_{\phi_1,\phi_2\in C^0(M;\mathbb{Z}_d)}g^k(M) \\
        &\qquad\times\exp\left[\frac{2\pi i}{d}\int \left(r\phi_1+\frac{ds}{N}\right)c - \left(\phi_2-\frac{dt}{N}\right)b - \phi_1\delta\phi_2\right],
    \end{split}
\end{equation}
where $\eta_b,\eta_c$ are Wilson lines of $b$ and $c$, respectively. When $s,t$ are multiples of $N/d$, the right hand side of \eqref{eq:N_k,s,t} can be transformed as
\begin{equation}
    \begin{split}
        \calN_k(M)&\eta_{b}^s(M)\eta_c^t(M) \\
        &= \frac{1}{|C^0(M;\mathbb{Z}_d)|}\sum_{\widetilde{\phi}_1,\widetilde{\phi}_2\in C^0(M;\mathbb{Z}_d)}g^k(M)\exp\left[\frac{2\pi i}{d}\int r\widetilde{\phi}_1c-\widetilde{\phi}_2b-\widetilde{\phi}_1\delta\widetilde{\phi}_2\right] \\
        &= \calN_k(M)
    \end{split}
\end{equation}
by performing the shift $\phi_1 \to \phi_1-r^{-1}\frac{ds}{N},\ \phi_2 \to \phi_2+\frac{dt}{N}$. Therefore, the labels are identified under the equivalence relations $s\sim s+\frac{N}{d}$ and $t\sim t+\frac{N}{d}$. We can define $\left(\frac{N}{d}\right)^2$ distinct non-invertible defects with grading $k$ by
\begin{equation}
    \calN_k^{s,t}(M) = \calN_k(M)\eta_{b}^s(M)\eta_c^t(M),
\end{equation}
where $s,t \in \mathbb{Z}_{N/d}$\footnote{We can also define them by multiplying $\eta_b^s\eta_c^t$ from the left since $\eta_b,\eta_c,\calN_k$ are all commutative.}.

Next, we compute the fusion rules $\calN_k\times \calN_{k'}$ between two non-invertible defects. Fusion obeys the group operation of $\mathbb{Z}_N$, and thus the grading of $\calN_k\times\calN_{k'}$ is $k+k'$. Let $d,d',D$ be the quantum dimensions of $\calN_k, \calN_{k'},\calN_{k+k'}$, respectively. Namely,
\begin{equation}
    d = \frac{N}{\gcd(N,k)},\quad d' = \frac{N}{\gcd(N,k')}, \quad D = \frac{N}{\gcd(N,k+k')}.
\end{equation}
Then, the fusion rule of $\calN_k\times\calN_{k'}$ is given by
\begin{equation}\label{eq:general fusion rule}
    \calN_k\times\calN_{k'} = \frac{dd'D}{l^2}\left(\sum_{i,j=0}^{l/D-1}\eta_b^{\frac{N}{l}i}\eta_c^{\frac{N}{l}j}\right)\calN_{k+k'},
\end{equation}
where $l=\mathrm{lcm}(d,d')$. We should note that when $k+k'=0\ \mathrm{mod}\ N$, we regard $\calN_{k+k'}$ as the identity. We can easily check the quantum dimension of the both side coincide.

Below, we derive these fusion rules by two complementary methods: the direct calculation and the representation-theoretic methods. The first method is difficult to apply to general $N$, and here we focus on the cases where $N$ is a prime or a product of distinct primes. This method is useful as it can also be employed when extending to general $N$-ality categories. The second method, while only applicable to $N$-ality categories that are Rep-categories like the one we consider here, allows for the computation of fusion rules for all $N$.


\subsubsection{$N=p$}
The simplest case is when $N$
is a prime number $p$. In this case, the non-invertible defect with grading $k\in\{1,2,\cdots,p-1\}$ is given by 
\begin{equation}\label{eq:non-inv. defect for N=p}
    \calN_k(M) = \frac{1}{|C^0(M;\mathbb{Z}_p)|}\sum_{\phi_1,\phi_2\in C^0(M;\mathbb{Z}_p)}g^k(M)\exp\left[\frac{2\pi i}{p}\int k\phi_1c - \phi_2b - \phi_1\delta\phi_2\right]
\end{equation}
and the quantum dimension of this defect is $p$. The fusion rules between two non-invertible defects \eqref{eq:general fusion rule} for the case where $N=p$ are given by 
\begin{equation} \label{fusion}
    \calN_k\times\calN_{k'}=
    \begin{cases*}
        \sum_{i,j=0}^{p-1}\eta_b^i\eta_c^j  & $(k+k'=p)$ \\
        p\calN_{k+k'} & (otherwise).
    \end{cases*}
\end{equation}

Let us check these fusion rules by direct calculation. From the expression of non-invertible defects \eqref{eq:non-inv. defect for N=p}, $\calN_k\times\calN_{k'}$ can be written as
\begin{equation}\label{eq:N_k*N_k' for N=p}
    \begin{split}
        \calN_k\times\calN_{k'} = & \frac{1}{|C^0(M;\mathbb{Z}_p)|^2}  \sum_{\substack{\phi_1,\phi_2,\phi_1',\phi_2'\\ \in C^0(M;\mathbb{Z}_p)}} g^{k+k'}(M) \\
        & \qquad \times\exp\left[\frac{2\pi i}{p}\int (k\phi_1+k'\phi_1')c - (\phi_2+\phi_2')b - \phi_1\delta\phi_2-\phi_1'\delta\phi_2'\right].
    \end{split}
\end{equation}
When $k+k'\not\equiv0\ \mathrm{mod}\ p$, introducing new variables, denoted as $\tilde{\phi}_1 = (k+k')^{-1}(k\phi_1+k'\phi_1'), \tilde{\phi}_2 = \phi_2 + \phi_2'$, we obtain
\begin{equation}
    \begin{split}
        \calN_k\times\calN_{k'} = & \frac{1}{|C^0(M;\mathbb{Z}_p)|^2}  \sum_{\substack{\phi_1,\phi_2,\phi_1',\phi_2'\\ \in C^0(M;\mathbb{Z}_p)}} g^{k+k'}(M) \exp\left[\frac{2\pi i}{p}\int (k+k')\tilde{\phi}_1c - \tilde{\phi}_2b -  \tilde{\phi}_1\delta\tilde{\phi}_2\right] \\
        & \qquad\times\exp\left[-\frac{2\pi i}{p}\int k'^{-1}(k+k')\Big(\phi_1-\tilde{\phi}_1\Big)\delta\Big(\phi_2-(k+k')^{-1}k\tilde{\phi}_2\Big)\right] \\
        = & \frac{1}{|C^0(M;\mathbb{Z}_p)|}\sum_{\Phi_1,\Phi_2\in C^0(M;\mathbb{Z}_p)}\exp\left[\frac{2\pi i}{p}\int \Phi_1\delta\Phi_2\right] \times \calN_{k+k'},
    \end{split}
\end{equation}
where we introduce new variables, denoted as $\Phi_1=-k'^{-1}(k+k')(\phi_1-\tilde{\phi}_1), \Phi_2=\phi_2-(k+k')^{-1}k\tilde{\phi}_2$ in the last equality. Summing over $\Phi_1$ introduces the following delta function constraint
\begin{equation}
    \frac{1}{|C^0(M;\mathbb{Z}_p)|}\sum_{\Phi_1\in C^0(M;\mathbb{Z}_p)}\exp\left[\frac{2\pi i}{p}\int \Phi_1\delta\Phi_2\right] = \delta(\delta\Phi_2).
\end{equation}
It implies that $\Phi_2$ must be a constant function, and thus we obtain
\begin{equation}
    \calN_k\times\calN_{k'} = p\calN_{k+k'}.
\end{equation}
When $k+k'\equiv 0\ \mathrm{mod}\ p$, we introduce new variables, denoted as $\tilde{\phi}_1 = \phi_1-\phi_1', \tilde{\phi}_2 = \phi_2+\phi_2'$ and \eqref{eq:N_k*N_k' for N=p} can be transformed as
\begin{equation}
    \calN_k\times\calN_{k'} = \frac{1}{|C^0(M;\mathbb{Z}_p)|^2}  \sum_{\substack{\phi_1,\phi_2,\phi_1',\phi_2'\\ \in C^0(M;\mathbb{Z}_p)}}\exp\left[\frac{2\pi i}{p}\int k\tilde{\phi}_1c - \tilde{\phi}_2b +\tilde{\phi}_1\delta\tilde{\phi}_2 - \phi_1\delta\tilde{\phi}_2 - \tilde{\phi}_1\delta\phi_2\right].
\end{equation}
Summing over $\phi_1,\phi_2$ enforces $\tilde{\phi}_1,\tilde{\phi}_2$ to be constant. Therefore, the term $\tilde{\phi}_1\delta\tilde{\phi}_2$ vanishes and we obtain
\begin{equation}
    \calN_k\times\calN_{k'} = \sum_{i,j=0}^{p-1}\eta_b^i\eta_c^j.
\end{equation}


\subsubsection{$N=pq$}
Let us consider the case of $N=pq$, where $p,q$ are two distinct prime numbers. In this case, non-invertible defects are  given by
\begin{align}
    \calN_{qr_1}(M) &= \frac{1}{|C^0(M;\mathbb{Z}_p)|}\sum_{\phi_1,\phi_2\in C^0(M;\mathbb{Z}_p)}g(M)^{qr_1}\exp\left[\frac{2\pi i}{p}\int r_1\phi_1c - \phi_2b - \phi_1\delta\phi_2\right],\label{N_qr}\\
    \calN_{pr_2}(M) &= \frac{1}{|C^0(M;\mathbb{Z}_q)|}\sum_{\phi_1,\phi_2\in C^0(M;\mathbb{Z}_q)}g(M)^{pr_2}\exp\left[\frac{2\pi i}{q}\int r_2\phi_1c - \phi_2b - \phi_1\delta\phi_2\right],\label{N_pr}\\
    \calN_{r_3}(M) &= \frac{1}{|C^0(M;\mathbb{Z}_{N})|}\sum_{\phi_1,\phi_2\in C^0(M;\mathbb{Z}_{N})}g(M)^{r_3}\exp\left[\frac{2\pi i}{N}\int r_3\phi_1c - \phi_2b - \phi_1\delta\phi_2\right]\label{N_r}.
\end{align}
where $r_1,r_2,r_3$ are coprime to $p,q,pq$, respectively. We first compute the product of \eqref{N_qr} and \eqref{N_pr}. From the expression \eqref{eq:general fusion rule}, it is expected that
\begin{equation}
    \calN_{qr_1}\times\calN_{pr_2} = \calN_{qr_1+pr_2}.
\end{equation}
Let us check this by explicit calculation.
\begin{equation}\label{eq:N_qr*N_pr}
    \begin{split}
        \calN_{qr_1}\times\calN_{pr_2} &= \frac{1}{|C^0(M;\mathbb{Z}_p)||C^0(M;\mathbb{Z}_q)|} \sum_{\substack{\psi_1,\psi_2\in C^0(M;\mathbb{Z}_p)\\ \chi_1,\chi_2\in C^0(M;\mathbb{Z}_q)}} g(M)^{qr_1+pr_2} \\
        & \qquad\times\exp\left[\frac{2\pi i}{N}\int (qr_1\psi_1+pr_2\chi_1)c - (q\psi_2+p\chi_2)b - q\psi_1\delta\psi_2-p\chi_1\delta\chi_2\right].
    \end{split}
\end{equation}
The third term and the forth term in the integrand can be transformed as
\begin{equation}
    \begin{split}
        q\psi_1\delta\psi_2 + p\chi_1\delta\chi_2 &= (qr_1+pr_2)^{-1}(qr_1\psi_1 + pr_2\chi_1)\delta(q\psi_2 + p\chi_2).
    \end{split}
\end{equation}
Then, we introduce new variables, denoted as $\phi_1=(qr_1+pr_2)^{-1}(qr_1\psi_1+pr_2\chi_1), \phi_2 = q\psi_2+p\chi_2$. We should note that $qr_1+pr_2$ is coprime to $N$ and $\phi_1,\phi_2$ can be regarded as $\mathbb{Z}_N$-valued cochains rather than $\mathbb{Z}_p$ or $\mathbb{Z}_q$-valued cochains. In terms of new variables,  we can rewrite \eqref{eq:N_qr*N_pr} as
\begin{equation}
    \begin{split}
        &\calN_{qr_1}\times\calN_{pr_2} \\
        &\quad= \frac{1}{|C^0(M;\mathbb{Z}_N)|} \sum_{\phi_1,\phi_2\in C^0(M;\mathbb{Z}_{N})}g^{qr_1+pr_2}(M)
        \times\exp\left[\frac{2\pi i}{N}\int (qr_1+pr_2)\phi_1c - \phi_2b - \phi_1\delta\phi_2\right] \\
        &\quad= \calN_{qr_1+pr_2}.
    \end{split}
\end{equation}

We can compute $\calN_{qr_1}\times\calN_{qr_1'}$ in the same way as the case $N=p$ and obtain
\begin{equation}
    \calN_{qr_1}\times\calN_{qr_1'} =
    \begin{cases*}
        \sum_{i,j=0}^{p-1}\eta_b^{qi}\eta_c^{qj}  & $(r_1+r_1'=p)$ \\
        p\calN_{q(r_1+r_1')} & (otherwise).
    \end{cases*}
\end{equation}

Other fusion rules can be obtained by the combination of the above results. For instance, we can compute $\calN_{r_3}\times\calN_{qr_1}$ as
\begin{equation}
    \calN_{r_3}\times\calN_{qr_1} = \calN_{pr_2'}\calN_{qr_1'}\calN_{qr_1} = 
    \begin{cases*}
        \sum_{i,j=0}^{p-1}\eta_b^{qi}\eta_c^{qj}\calN_{pr_2'}  & $(r_1+r_1'=p)$ \\
        p\calN_{r_3+qr_1} & (otherwise),
    \end{cases*}
\end{equation}
where $r_3 = qr_1'+pr_2'$\footnote{Note that all defects are commutative.}.

\subsubsection{From the perspective of representation theory}
The $N$-ality category we are discussing can actually be regarded as the Rep-category of a certain non-Abelian group. In this section, we use this perspective to derive the fusion rules for general $N$.

Although the $N$-ality category we discuss is obtained by gauging $\mathbb{Z}_N^{\rmB}\times\mathbb{Z}_N^{\rmC}$ subgroup of $\mathbb{Z}_N^{\rmA}\times\mathbb{Z}_N^{\rmB}\times\mathbb{Z}_N^{\rmC}$ with a type III anomaly, we obtain the following non-Abelian group when gauging $\mathbb{Z}_N^{\rmA}$\footnote{As studied in \cite{Tachikawa:2017gyf}, the symmetry $G'$ obtained by gauging $\mathbb{Z}_N^{\rmA}$ is described by the extension $\mathbb{Z}_N^{\rmA}\rightarrow G' \rightarrow \mathbb{Z}_N^{\rmB}\times\mathbb{Z}_N^{\rmC}$ with a 2-cocycle $e((b_1,c_1),(b_2,c_2))\coloneqq b_1c_2$. We can check $G'\simeq G$ by computing the multiplication rule of each group explicitly.
}:
\begin{equation}
    G = \left( \mathbb{Z}_N^{\rmA}\times\mathbb{Z}_N^{\rmB} \right) \rtimes_{\rho} \mathbb{Z}_N^{\rmC},
\end{equation}
where $\rho$ is
\begin{equation}
    \rho:\mathbb{Z}_N^{\rmC}\rightarrow \mathrm{Aut}(\mathbb{Z}_N^{\rmA}\times\mathbb{Z}_N^{\rmB}),\quad \rho(c)(a,b)=(a-cb,b).
\end{equation}
This non-Abelian group is knwon as the Heisenberg group over $\mathbb{Z}_N$ for a prime $N$. Gauging the entire group $G$ is equivalent to gauging $\mathbb{Z}_N^{\rmB}\times\mathbb{Z}_N^{\rmC}$ subgroup of $\mathbb{Z}_N^{\rmA}\times\mathbb{Z}_N^{\rmB}\times\mathbb{Z}_N^{\rmC}$ with a type III anomaly, and as a result, the $N$-ality symmetry is obtained by gauging $G$. This implies that the $N$-ality symmetry is equivalent to the $\Rep(G)$ symmetry.

Therefore, the defects in the $N$-ality symmetry can be regarded as Wilson lines, or equivalently characters of representations of $G$. As stated in Appendix \ref{appendix A}, an irreducible representation of $G$ is labeled by a divisor $d$ of $N$, $r$ coprime to $d$, and $s,t\in\mathbb{Z}_{N/d}$. The non-invertible defect $\calN_k$ corresponds to the irreducible representation with $d = N/\gcd(N,k), r = dk/N, s=t=0$\footnote{We can easily check the quantum dimension and the grading of them are same.}.

From the expression \eqref{eq:caharacter of irrep}, one can see that the defect $\calN_k$, as the functional of gauge fields, does not depend on $b,c$. We can check this explicitly from the expression of defects \eqref{eq:general N_k}. In \eqref{eq:general N_k}, summing over $\phi_1\in C^0(M;\mathbb{Z}_d)$ imposes the condition $c = r^{-1}\delta\phi_2$, and this implies that the period of $c$ is restricted to multiples of $d$. When this condition holds, we can set $c$ to 0 by using the gauge transformation \eqref{eq:gauge trfm. for Z_d}. The same argument is true for $b$, and thus $\calN_k$ is only nonzero when the period of $b$ is a multiple of $d$ and we can set $b$ to 0. As a result, when the periods of $b,c$ are multiples of $d$, the defect $\calN_k$ can be written as
\begin{equation}
    \begin{split}
        \calN_k(M) &= \frac{1}{|C^0(M;\mathbb{Z}_d)|}\sum_{\phi_1,\phi_2\in C^0(M;\mathbb{Z}_d)}g^k(M)\exp\left[-\frac{2\pi i}{d}\int \phi_1\delta\phi_2\right] \\
        &= d\times g^k(M).
    \end{split}
\end{equation}
In summary,
\begin{equation}
    \calN_k(M) = 
    \begin{cases*}
        d\times g^k(M) & ($\int b,\int c\equiv0\ \mathrm{mod}\ d$) \\
        0 & (otherwise).
    \end{cases*}
\end{equation}

We can prove the fusion rules \eqref{eq:general fusion rule} with this expression. Substituting this expression for the right hand side of \eqref{eq:general fusion rule}, we obtain
\begin{equation}\label{eq:RHS of general fusion rule}
    \frac{dd'D^2}{l^2}\left(\sum_{i,j=0}^{l/D-1}\eta_b^{\frac{N}{l}i}\eta_c^{\frac{N}{l}j}\right) g^{k+k'}(M)
\end{equation}
when $\int b,\int c\equiv0\ \mathrm{mod}\ D$. We assume $\int b = Dx,\int c = Dy$ with $x,y\in\mathbb{Z}_{N/D}$, then $\eta_b^{N/l}=e^{2\pi iDx/l},\eta_c^{N/l}=e^{2\pi iDy/l}$ and thus the sum in \eqref{eq:RHS of general fusion rule} is nonzero if and only if $x,y$ are multiples of $\frac{l}{D}$. These conditions are equivalent to conditions $\int b, \int c \equiv 0 \ \mathrm{mod}\ l$. When these conditions hold, the right hand side of \eqref{eq:general fusion rule} can be written as
\begin{equation}
     \frac{dd'D^2}{l^2}\left(\sum_{i,j=0}^{l/D-1}\eta_b^{\frac{N}{l}i}\eta_c^{\frac{N}{l}j}\right) g^{k+k'}(M) = dd'\times g^{k+k'}(M).
\end{equation}
This is, indeed, equivalent to the left hand side of \eqref{eq:general fusion rule} since $l=\mathrm{lcm}(d,d')$.


\section{Non-invertible SPT phases} \label{section 3}
In this section, we classify the SPT phases for the $N$-ality symmetry constructed in Section \refeq{section 2}. Furthermore, we construct lattice Hamiltonians which realize some of them.

\subsection{Correspondence between symmetric gapped phases} 
To study symmetric gapped phases without referring to the underlying dynamics of the theory, the SymTFT framework is extremely useful~\cite{Ji:2019jhk,Gaiotto:2020iye,Apruzzi:2021nmk,Kong:2020cie,Freed:2022qnc,Freed:2022iao}. Applications of SymTFT are, for example, studied in~\cite{Kaidi:2022cpf,Kaidi:2023maf,Burbano:2021loy,vanBeest:2022fss,Chen:2023qnv,Sun:2023xxv,Apruzzi:2023uma,Cao:2023rrb,Duan:2023ykn,Antinucci:2024zjp,Brennan:2024fgj,Bhardwaj:2024ydc}. The SymTFT for $(1+1)$-dimensional theory with a fusion category symmetry $\mathcal{C}$ is a $(2+1)$-dimensional TQFT with two boundaries. The first one is the topological boundary, which encodes the data of the fusion category symmetry $\mathcal{C}$. The second one is the physical boundary, where the original $(1+1)$-dimensional theory lives and is not necessarily topological. Since the bulk is topological, one can shrink it and obtain the original $(1+1)$-dimensional theory with a fusion category symmetry $\mathcal{C}$. Replacing the topological boundary condition with another one, the distinct fusion category symmetry, which relates with $\mathcal{C}$ by a topological manipulation, is realized.
Topological boundary conditions of the SymTFT are in one-to-one correspondence with Lagrangian algebras of Drinfeld center $\mathcal{Z}(\mathcal{C})$ of $\mathcal{C}$\footnote{$\mathcal{Z}(\mathcal{C})$ is a modular tensor category and captures the anyon data of the TQFT.} (see \cite{Lan:2014uaa,Kaidi:2021gbs,Kobayashi:2022vgz,Davydov:2010kfz,Fuchs:2012dt,Davydov:2011dqx} for reviews). A Lagrangian algebra specifies the topological lines in $\mathcal{Z}(\mathcal{C})$ that condense on the topological boundary.
In the SymTFT perspective, topological manipulations are interpreted as the action of 0-form symmetries in the bulk SymTFT on the topological boundaries.

Next, we explain how we can classify symmetric gapped phases with $\mathcal{C}$ in (1+1)-dimensions by using SymTFT~\cite{Thorngren:2019iar,Bhardwaj:2023idu, Bhardwaj:2023fca,Bhardwaj:2024qiv,Bhardwaj:2025piv}. To obtain gapped phases, one also takes the physical boundary to be topological. We fix the topological boundary to $\mathcal{A}_{\mathcal{C}}$, which is a Lagrangian algebra corresponding to the symmetry $\mathcal{C}$. We denote the Lagrangian algebra chosen on the physical boundary by $\mathcal{A}_{\mathrm{phys}}$. Shrinking the bulk, we obtain the $\mathcal{C}$-symmetric gapped phase corresponding to $\mathcal{A}_{\mathrm{phys}}$. If we perform a topological manipulation, we obtain the different topological boundary $\mathcal{A}_{\mathcal{C}'}$, and obtain the $\mathcal{C}'$-symmetric gapped phase when shrinking the bulk. Thus, if two fusion category symmetries $\mathcal{C}$ and $\mathcal{C}'$ can be connected by a certain topological manipulation\footnote{Such a topological manipulation exists if $\mathcal{Z}(\mathcal{C})\simeq\mathcal{Z}(\mathcal{C}')$, i.e., $\mathcal{C}$ and $\mathcal{C}'$ are Morita equivalent.}, there is the following one-to-one correspondence:
\begin{align}
    \{\mathcal{C}\text{-\textit{symmetric gapped phases}}\}\xleftrightarrow{\text{1\ :\ 1\ }}\{\mathcal{C}'\text{-\textit{symmetric gapped phases}}\}. 
\end{align}


\subsection{Classification of non-invertible SPT phases}
We consider the correspondence given by
\begin{align}   \label{correspondence}\underbrace{\mathbb{Z}_N^{\rmA}\times\mathbb{Z}_N^{\rmB}\times\mathbb{Z}_N^{\rmC}}_{\text{type III anomaly}}
\xrightarrow[\textbf{S}]{\text{gauge}}\ \mathbb{Z}_N^{\rmA}\text{-graded fusion category},
\end{align}
where \textbf{S} is the gauging of $\mathbb{Z}_N^{\rmB}\times\mathbb{Z}_N^{\rmC}$ symmetry. As mentioned above, there is a one-to-one correspondence between the gapped phases associated with each symmetry. In this subsection, we classify the SPT phases for the $N$-ality symmetry by using this correspondence. More precisely, we derive the conditions for a $\mathbb{Z}_N^{\rmA}\times\mathbb{Z}_N^{\rmB}\times\mathbb{Z}_N^{\rmC}$-symmetric gapped phase to be mapped to an SPT phase of the $N$-ality symmetry by the gauging operation $\mathbf{S}$.

We note that SPT phases of the $N$-ality symmetry can be regarded as SPT phases of $\widehat{\mathbb{Z}}_N^{\rmB}\times\widehat{\mathbb{Z}}_N^{\rmC}$, which is a subsymmetry of the $N$-ality symmetry. Thus, we first ignore the $\mathbb{Z}_N^{\rmA}$ symmetry and derive the conditions for a $\mathbb{Z}_N^{\rmB}\times\mathbb{Z}_N^{\rmC}$-symmetric gapped phase to be mapped to a level-$k$ SPT phase of $\widehat{\mathbb{Z}}_N^{\rmB} \times \widehat{\mathbb{Z}}_N^{\rmC}$ symmetry by the gauging operation $\mathbf{S}$\footnote{$\mathbb{Z}_N\times\mathbb{Z}_N$-SPT phases are classified by $H^2(\mathbb{Z}_N\times\mathbb{Z}_N;U(1))\simeq\mathbb{Z}_N$.}. Then, we recall the $\mathbb{Z}_N^{\rmA}$ symmetry and derive the conditions for $\mathbb{Z}_N^{\rmA}\times\mathbb{Z}_N^{\rmB}\times\mathbb{Z}_N^{\rmC}$-symmetric gapped phases.

The partition function of the level-$k$ SPT phase of $\widehat{\mathbb{Z}}_N^{\rmB} \times \widehat{\mathbb{Z}}_N^{\rmC}$ symmetry is given by
\begin{equation}
    Z_k[B,C]=\exp[\frac{2\pi i}{N}k\int B\cup C],
\end{equation}
where $B,C$ are background gauge fields for $\widehat{\mathbb{Z}}_N^{\rmB}, \widehat{\mathbb{Z}}_N^{\rmC}$, respectively. By implementing $\mathbf{S}^{-1}$ on a level-$k$ SPT\footnote{Note that $\mathbf{S}^2=1$ and thus the operation $\mathbf{S}^{-1}$ is equivalent to the operation $\mathbf{S}$.}, we obtain
\begin{equation}\label{eq:S of level k SPT}
    \begin{split}
        \frac{1}{|H^1(X;\mathbb{Z}_N)|}&\sum_{b,c\in H^1(X;\mathbb{Z}_N)}
        \exp[\frac{2\pi i}{N}\int kb\cup c + b\cup C + c\cup B] \\
        &\qquad=|H^1(X;\mathbb{Z}_d)|\ \delta^{(N)}(xB)\delta^{(N)}(xC)\exp[\frac{2\pi i}{x} y^{-1}\int \frac{B}{d}\cup \frac{C}{d}],
    \end{split}
\end{equation}
where
\begin{equation}
    N=dx,\quad k=dy,\quad \text{gcd}(N,k)=d,\quad \text{gcd}(x,y)=1.
\end{equation}
The derivation of \eqref{eq:S of level k SPT} is given in Appendix \ref{appendix D}. The partition function \eqref{eq:S of level k SPT} describes the phase where $\mathbb{Z}_N^{\rmB}\times\mathbb{Z}_N^{\rmC}$ is broken to $\mathbb{Z}_x^{B}\times\mathbb{Z}_x^{C}$ and equipped with a level-$y^{-1}$ $\mathbb{Z}_x^{B}\times\mathbb{Z}_x^{C}$ SPT phase \cite{Thorngren:2019iar}.

Next, we recall $\mathbb{Z}_N^{\rmA}$ symmetry and derive the conditions for $\mathbb{Z}_N^{\rmA}\times\mathbb{Z}_N^{\rmB}\times\mathbb{Z}_N^{\rmC}$-symmetric gapped phases. Of course, not only defects of $\widehat{\mathbb{Z}}_N^{\rmB} \times \widehat{\mathbb{Z}}_N^{\rmC}$ symmetry, but also non-invertible symmetry defects must be unbroken in an SPT phase associated with the $N$-ality symmetry. Thus, the symmetry defects corresponding to the non-invertible symmetry defects $\calN_1,\cdots,\calN_{N-1}$ must be unbroken in the pregauged theory. This condition is equivalent to that $\mathbb{Z}_N$ symmetry generated by $(1,m,n)\in\mathbb{Z}_N^{\rmA}\times\mathbb{Z}_N^{\rmB}\times\mathbb{Z}_N^{\rmC}$ is unbroken\footnote{In section \ref{sec 2.3}, we constructed non-invertible symmetry defects $\calN_1$ from $(1,0,0)$ line. However, $\mathbb{Z}_N^{\rmB}\times\mathbb{Z}_N^{\rmC}$ lines are absorbed when gauging and thus $(1,m,n)$ line also becomes $\calN_1$ by gauging.}. Therefore, the unbroken symmetry $K\subset\mathbb{Z}_N^{\rmA}\times\mathbb{Z}_N^{\rmB}\times\mathbb{Z}_N^{\rmC}$ in the pregauged theory must be
\begin{equation}
    K=\langle(1,m,n), (0,d,0), (0,0,d)\rangle\simeq\mathbb{Z}_N\times\mathbb{Z}_x^{\rmB}\times\mathbb{Z}_x^{\rmC} \subset \mathbb{Z}_N^{\rmA}\times\mathbb{Z}_N^{\rmB}\times\mathbb{Z}_N^{\rmC},\quad m,n\in\mathbb{Z}_d. \label{K}
\end{equation}
{Note that since $K$ contains the element $(0,d,0)$, the label $m$ is identified with $m+d$ (and similarly for n).

However, not all subgroups of this form can serve as unbroken subgroups. The unbroken symmetry $K$ must be free of type III anomalies. We represent the gauge field for $K$ as $(\tilde{A},\tilde{B},\tilde{C})$ and the relation with the original gauge field $(A,B,C)$ is given by
\begin{equation}
    A = \tilde{A},\qquad B = m\tilde{A}+d\tilde{B},\qquad C = n\tilde{A}+d\tilde{C}.
\end{equation}
Then, the type III anomaly \eqref{type III anomaly} is represented in terms of the gauge field for $K$ as
\begin{equation}
    \exp[\frac{2\pi i}{N}\qty(mn\int \tilde{A}^3 + md\int \tilde{A}^2\cup \tilde{B} + nd\int \tilde{A}^2\cup \tilde{C} + d^2\int \tilde{A}\cup \tilde{B}\cup \tilde{C})].
\end{equation}
Using the fact we mentioned in the Appendix \ref{appendix D}, the anomaly free condition is given by
\begin{itemize}
    \item odd $N$ 
    \begin{equation}
        d^2\equiv 0 \mod N
    \end{equation}

    \item even $N$ 
    \begin{equation}
         d^2\equiv 0 \mod N,\quad \text{and $mn$ is even.}
    \end{equation}
\end{itemize}

Thus, when $N$ is odd, the number of gapped phases satisfying the above conditions is
\begin{equation}
    \sum_{k}\Big(\mathrm{gcd}(N,k)\Big)^2, \label{odd}
\end{equation}
and when $N$ is even, it is
\begin{equation}
    \sum_{k}\frac{3}{4}\Big(\mathrm{gcd}(N,k)\Big)^2, \label{even}
\end{equation}
where the sum runs over $k$ such that $\left(\mathrm{gcd}(N,k)\right)^2\equiv 0 \mod N$. This completes the classification of SPT phases associated with the $N$-ality symmetry.

One may wonder whether we can obtain other gapped phases of $\mathbb{Z}_N^{\rmA}\times\mathbb{Z}_N^{\rmB}\times\mathbb{Z}_N^{\rmC}$ symmetry by stacking SPT phases associated with the subgroups $\mathbb{Z}_N\times\mathbb{Z}_x^{\rmB},\mathbb{Z}_N\times\mathbb{Z}_x^{\rmC}\subset K$ symmetry. However, in fact, such SPT stacking operations do not generate new phases: they leave the phase unchanged. We explain this in detail in Appendix~\ref{appendix D}.

To summarize, the classification of the SPT phases for the $N$-ality symmetry is as follows:
\begin{itemize}
    \item For odd $N$, there are $\sum_{k}(\mathrm{gcd}(N,k))^2$ SPT phases.
    \item For even $N$, there are $\sum_{k}\frac{3}{4}(\mathrm{gcd}(N,k))^2$ SPT phases.
\end{itemize}
We obtain, for example, three distinct SPT phases in the case of $\Rep(D_8)$ (i.e. $N=2$ case), which was studied in \cite{Thorngren:2019iar, Seifnashri:2024dsd}. Mathematically, these results correspond to the classification of fiber functors of the $N$-ality category and are consistent with the mathematical classification of fiber functors \cite{Meir2011,Lu:2025gpt,Ostrik:2002ohv}. 

Let us explain a bit more detail the relation to the mathematical classification of fiber functors. Our setup is as follows. We consider a group-theoretical fusion category $\mathcal{C}(G,\omega,H,\psi)$, which is the dual symmetry obtained by gauging a non-anomalous subgroup $H\subset G$ with discrete torsion $\psi\in H^2(H;U(1))$, where $G$ has an anomaly characterized by $\omega\in H^3(G;U(1))$. As we have discussed, there is a one-to-one correspondence between gapped phases of $G$ with anomaly $\omega$ and gapped phases of $\mathcal{C}(G,\omega,H,\psi)$. The gapped phase of $G$ with anomaly $\omega$ can be characterized by the data $(K,\psi_K)$, where $K$ is the non-anomalous unbroken subgroup of $G$ and $\psi_K \in H^2(K;U(1))$ is the SPT phase of $K$\footnote{Note that we must take into account the equivalence relation between two gapped phases, as discussed in Appendix \ref{appendix D}.}. Then, a gapped phase of $G$ with anomaly is mapped to an SPT phase of $\mathcal{C}(G,\omega,H,\psi)$ by gauging $H$ with discrete torsion $\psi\in H^2(H;U(1))$ if and only if the following conditions hold:
\begin{itemize}
    \item $G=HK$
    \item the 2-cocycle $\frac{\psi_K|_{H\cap K}}{\psi|_{H\cap K}}\in Z^2(H\cap K,U(1))$ is non-degenerate.
\end{itemize}
The case we focus on corresponds to $G=\mathbb{Z}_N^{\mathrm{A}} \times\mathbb{Z}_N^{\mathrm{B}} \times\mathbb{Z}_N^{\mathrm{C}}$, $H=\mathbb{Z}_N^{\mathrm{B}} \times\mathbb{Z}_N^{\mathrm{C}}$, $\psi=1$, and $\omega$ being a type III anomaly. We can verify that the conditions for realizing the SPT phase obtained from the physical argument above agree with those required for the mathematical fiber functor.


\subsection{Lattice Hamiltonians}
In this subsection, we construct lattice Hamiltonians for some of SPT phases associated with the $N$-ality symmetry. In particular, we focus on the unbroken subgroup $K=\langle(1,m,0)\rangle$ and $K=\langle(1,0,m)\rangle$.
 Here, unlike in the previous section, we construct the N-ality symmetry not by just the gauging (\textbf{S}) but by the twisted gauging (\textbf{TST}). Note that the N-ality symmetry obtained by TST is the same as the one in the previous section\footnote{The only difference between the two constructions is the stacking of an SPT (\textbf{T}), which does not change the symmetry of the theory.}. In this case, the N-ality symmetry is generated by the topological manipulation \textbf{TSTST}$^{-1}$.

We consider an anomalous $\mathbb{Z}_N^{\rm V}\times \mathbb{Z}_N^{\rm e}\times\mathbb{Z}_N^{\rm o}$ symmetry generated by\footnote{In the previous discussion, the indices V, e, and o correspond to A, B, and C, respectively.}
\begin{equation} \label{sym_op}
     V=\prod_{n=1}^{L/2}CZ_{2n-1,2n}CZ_{2n,2n+1}^\dagger, \quad\eta_e=\prod_{j:\text{even}}X_j,\quad \eta_o=\prod_{j:\text{odd}}X_j,
\end{equation}
where $X$ and $Z$ are the $\mathbb{Z}_N$ shift and clock operator, respectively, and $CZ$ is a controlled-$Z$ gate for $\mathbb{Z}_N$, (see Appendix \refeq{appendix B} for details). Here, we assume that the system is defined on a periodic chain with an even number of sites $L$. These symmetry operators have a type III anomaly, (see Appendix \refeq{appendix C}). The topological manipulation \textbf{S} and \textbf{T} on the lattice can be realized as following transformations 
\begin{equation}
    \textbf{S}:X_j\rightsquigarrow Z_{j-1}^\dagger Z_{j+1},\quad Z_{j-1}^\dagger Z_{j+1}\rightsquigarrow X_j ,
\end{equation}
\begin{equation}
\begin{aligned}
   \textbf{T}: X_{2n}\rightarrow Z_{2n-1}X_{2n}Z_{2n+1}^\dagger,\quad
    X_{2n+1}\rightarrow Z_{2n}^\dagger X_{2n+1}Z_{2n+2},\quad Z_{j}\rightarrow Z_{j}.
\end{aligned}
\end{equation}
See Appendix \refeq{appendix B} for details.
\begin{enumerate}[label=(\roman*)]
\item $\mathbb{Z}_N^{\rm V}$ preserving phase \\[2mm]
The simplest case of SSB patterns which satisfy the above conditions is the $\mathbb{Z}_N^{\rm V}$ preserving phase given by
\begin{equation} \label{SSB}
    H_{0}=-\sum_{n=1}^{L/2}(Z_{2n-1}Z_{2n+1}^{\dagger}+Z_{2n}Z_{2n+2}^{\dagger})+h.c..
\end{equation}
This Hamiltonian has a $\mathbb{Z}_N\times\mathbb{Z}_N\times\mathbb{Z}_N$ symmetry generated by \eqref{sym_op}. The ground states stabilize the $L-2$ generators
\begin{equation}
    Z_{2n-1}Z_{2n+1}^\dagger=1, \quad Z_{2n}Z_{2n+2}^\dagger=1,\quad \forall n.
\end{equation}
This leads to an $N^2$-fold degeneracy resulting from SSB of the $\mathbb{Z}_N^{\rm e}\times\mathbb{Z}_N^{\rm o}$ symmetry, and these ground states preserve the $\mathbb{Z}_N^{\rm V}$ symmetry.

Next we perform the twisted gauging of $\mathbb{Z}_N^{\rm e}\times\mathbb{Z}_N^{\rm o}$ (\textbf{TST}) to obtain the non-invertible SPT phase,
\begin{equation}
    \hat{H}_{\text{SPT}_1}=-\sum_{n=1}^{L/2}(Z_{2n-1}X_{2n}Z_{2n+1}^{-1}+Z_{2n}^{-1}X_{2n+1}Z_{2n+2})+h.c..
\end{equation}
This Hamiltonian is a level-$1$ SPT phase for the $\mathbb{Z}_N^{\rm e}\times\mathbb{Z}_N^{\rm o}$ symmetry \cite{Chen:2011pg,Tsui:2017ryj}, and thus it has a unique gapped ground state, denoted by $\ket{\text{SPT}_1}$, which is stabilized by the following $L$ generators, 
\begin{equation}
    Z_{2n-1}X_{2n}Z_{2n+1}^{-1}=1,\quad Z_{2n}^{-1}X_{2n+1}Z_{2n+2}=1,\quad \forall n.
\end{equation}
Furthermore, this Hamiltonian is invariant under the topological manipulation $(\textbf{TSTST}^{-1})$. Therefore, we conclude that this level-$1$ SPT phase is a non-invertible SPT phase.  
\item $\mathbb{Z}_N=\langle(1,m,0)\rangle\subset \mathbb{Z}_N^{\rm V} \times \mathbb{Z}_N^{\rm e}\times\mathbb{Z}_N^{\rm o}$ preserving phase,$\quad m\in\{1,2,\cdots,N-1\}$\\[2mm]
The Hamiltonian that describes this SSB pattern is given by
\begin{equation} \label{SSB_odd}
         H_{\text{odd},m}=-\sum_{n=1}^{L/2}\qty(\omega^{m}Z_{2n-1}Z_{2n+1}^\dagger+\sum_{k=0}^{N-1}Z_{2n-1}^{-k}Y_{2n}Z_{2n+1}^{2k}Y_{2n+2}^\dagger Z_{2n+3}^{-k})+h.c..
\end{equation}
where $Y=e^{\frac{N-1}{N}\pi i}X^\dagger Z$ is the generalization of Pauli matrix $\sigma^y$\footnote{$Y$ is a unitary operator and satisfies $Y^N=I$ where $I$ is an identity operator.}.
Here, we assume $L$ is a multiple of $2N$. 
This is a commuting projector Hamiltonian and the ground states stabilize the following $L-2$ generators,
\begin{equation}
    Z_{2n-1}Z_{2n+1}^\dagger=\omega^{-m},\quad Y_{2n}Y_{2n+2}^\dagger=1,\quad \forall n.
\end{equation}
Note that $Z_{2n-1}^\dagger Z_{2n+1}^2Z_{2n+3}^\dagger=1$ is automatically satisfied by $Z_{2n-1}Z_{2n+1}^\dagger=\omega^{-m}$. This leads to an $N^2$-fold degeneracy. One can check that these ground states preserve the $\mathbb{Z}_N=\langle(1,m,0)\rangle$ symmetry.

Let us perform the twisted gauging (\textbf{TST}), 
\begin{equation}
    \begin{aligned}
        \hat{H}_{\text{odd},m}&=-\sum_{n=1}^{L/2}\omega^{m}Z_{2n-1}X_{2n}Z_{2n+1}^{-1}
        -\sum_{n=1}^{L/2}\sum_{k=0}^{N-1}\chi_{2n+1}^{(k)}+h.c. ,\\
        \chi_{2n+1}^{(k)}&=Z_{2n-1}^{k+2}X_{2n}^{k+1}Z_{2n}Z_{2n+1}^{-k-2}X_{2n+1}^{-1}Z_{2n+1}^{-k-2}Z_{2n+2}^{-1}X_{2n+2}^{-k-1}Z_{2n+3}^{k+2}.
\end{aligned}
\end{equation}
This Hamiltonian has a unique gapped ground state, denoted by $\ket{\mathrm{odd_m}}$, stabilized by the following $L$ generators,
\begin{equation}
    Z_{2n-1}X_{2n}Z_{2n+1}^{-1}=\omega^{-m},\quad \chi_{2n+1}^{(0)}=1,\quad \forall n.
\end{equation}
Furthermore, this Hamiltonian is invariant under the topological manipulation $(\textbf{TSTST}^{-1})$. More concretely, the first sum of $\hat{H}_{\text{odd},m}$ is invariant, and $\chi^{(k)}$ is mapped to $\chi^{(k+1)}$ under the topological manipulation $(\textbf{TSTST}^{-1})$.

\item $\mathbb{Z}_N=\langle(1,0,m)\rangle\subset \mathbb{Z}_N^{\rm V} \times \mathbb{Z}_N^{\rm e}\times\mathbb{Z}_N^{\rm o}$ preserving phase,$\quad m\in\{1,2,\cdots,N-1\}$ \\[2mm]
Similar to (ii), the Hamiltonian that describes this SSB pattern is given by
\begin{equation} \label{SSB_even}
         H_{\text{even},m}=-\sum_{n=1}^{L/2}\qty(\omega^{-m}Z_{2n}Z_{2n+2}^\dagger+\sum_{k=0}^{N-1}Z_{2n-2}^kY_{2n-1}Z_{2n}^{-2k}Y_{2n+1}^\dagger Z_{2n+2}^k)+h.c..
\end{equation}
Here, we assume $L$ is a multiple of $2N$. This is also a commuting projector Hamiltonian and the ground states stabilize the following $L-2$ generators,
\begin{equation}
    Z_{2n}Z_{2n+2}^\dagger=\omega^{m},\quad Y_{2n-1}Y_{2n+1}^\dagger=1 ,\quad \forall n.
\end{equation}
Note that $Z_{2n-2} Z_{2n}^{-2}Z_{2n+2}=1$ is automatically satisfied by $Z_{2n}Z_{2n+2}^\dagger=\omega^{m}$. This leads to an $N^2$-fold degeneracy and these ground states preserve the $\mathbb{Z}_N=\langle(1,0,m)\rangle$ symmetry.

Let us perform the twisted gauging (\textbf{TST}), 
\begin{equation}
\begin{aligned}
        \hat{H}_{\text{even},m}&=-\sum_{n=1}^{L/2}\omega^{-m}Z_{2n}^{-1}X_{2n+1}Z_{2n+2}
        -\sum_{n=1}^{L/2}\sum_{k=0}^{N-1}\widetilde{\chi}_{2n}^{(k)}+h.c., \\
        \widetilde{\chi}_{2n}^{(k)}&=Z_{2n-2}^{k+2}X_{2n-1}^{-k-1}Z_{2n-1}^{-1}Z_{2n}^{-k-2}X_{2n}^{-1}Z_{2n}^{-k-2}Z_{2n+1}X_{2n+1}^{k+1}Z_{2n+2}^{k+2}.
\end{aligned}
\end{equation}
This Hamiltonian has a unique gapped ground state, denoted by $\ket{\mathrm{even_m}}$, stabilized by the following $L$ generators:
\begin{equation}
    Z_{2n}^{-1}X_{2n+1}Z_{2n+2}=\omega^{m},\quad \widetilde{\chi}_{2n}^{(0)}=1,\quad \forall n.
\end{equation}
Similarly to $\hat{H}_{\text{odd},m}$, one can check that this Hamiltonian is invariant under the topological manipulation $(\textbf{TSTST}^{-1})$.
\end{enumerate}
Thus, we have constructed $(2N-1)$ SPT phases for the $N$-ality symmetry, $\ket{\text{SPT}_1}$, $\ket{\text{odd}_m}$, and $\ket{\text{even}_m}$ where $m\in\{1,2,\cdots,N-1\}$. These SPT phases can be distinguished by different symmetry breaking patterns in dual theories.  In the case of $N=2$, this result corresponds to $\mathrm{Rep}(D_8)$ SPT phases constructed in \cite{Seifnashri:2024dsd}.\\
\\
Finally, we show that these non-invertible SPT phases are in the same phase as $\mathbb{Z}_N^{\rm e}\times\mathbb{Z}_N^{\rm o}$ SPT phases. To check this, we consider the interface between $\ket{\mathrm{SPT}_1}$ and $\ket{\text{odd}_m}$. The interface Hamiltonian that preserves the non-invertible symmetry can be defined as
\begin{equation}
    \begin{aligned}
        H_{\mathrm{SPT_1}|\mathrm{odd_m}}=&-\sum_{n=1}^{\ell/2}(Z_{2n-1}X_{2n}Z_{2n+1}^{-1}+Z_{2n-2}^{-1}X_{2n-1}Z_{2n})\\
        &-\sum_{n=\ell/2+1}^{L/2}\omega^mZ_{2n-1}X_{2n}Z_{2n+1}^{-1}-\sum_{n=\ell/2+1}^{L/2-2}\sum_{k=0}^{N-1}\chi_{2n+1}^{(k)}+h.c..
    \end{aligned}
\end{equation}
Here we consider a periodic chain of $L$ sites, where $\ket{\mathrm{SPT}_1}$ lives in the region between sites $1$ and $\ell$, and $\ket{\text{odd}_m}$ lives in the region between sites $\ell$ and $L$. We assume $L-\ell$ to be a multiple of $2N$. The ground states of this Hamiltonian are stabilized by the following $L-2$ generators,
\begin{equation}
    \begin{aligned}
        Z_{2n-1}X_{2n}Z_{2n+1}^{-1}&=1\quad &&\text{for}\ \  n=1, 2,\cdots, \ell/2,\\
        Z_{2n-2}^{-1}X_{2n-1}Z_{2n}&=1\quad &&\text{for}\ \ n=1, 2,\cdots, \ell/2,\\
        Z_{2n-1}X_{2n}Z_{2n+1}^{-1}&=\omega^{-m}\quad &&\text{for}\ \  n=\ell/2+1,\cdots, L/2,\\
        \chi_{2n+1}^{(k)}&=1\quad &&\text{for}\ \ n=\ell/2+1,\cdots, L/2-2.\\
    \end{aligned}
\end{equation}
Thus, the ground states are $N^2$-fold degenerate, which can be characterized by the edge modes at the interfaces. We next discuss the action of $\eta_e$ and $\eta_o$ on ground states $\ket{\psi}$. These actions can be written as 
\begin{equation}
    \begin{aligned}
        \eta_o\ket{\psi}&=\eta_o^L\eta_o^R\ket{\psi},\quad \eta_o^L=Z_{L-2}^{-1}X_{L-2}X_{L-1}Z_L,\quad \eta_o^R=Z_{\ell}^{-1}X_{\ell+1}X_{\ell+2}^{-1}Z_{\ell+2}\\
        \eta_e\ket{\psi}&=\ket{\psi},
    \end{aligned}
\end{equation}
where $L$ and $R$ are labels for two interfaces at sites $j=L$ and $j=\ell$, respectively. Since $\eta_e$ acts trivially on the ground states, no projective representation between $\eta_e$ and $\eta_o$ arises at the interface. This result indicates that $\ket{\mathrm{SPT}_1}$ and $\ket{\text{odd}_m}$ belong to the same phases as $\mathbb{Z}_N^{\rm e}\times\mathbb{Z}_N^{\rm o}$ SPT phases. Indeed, if we disregard the non-invertible symmetry, one can construct an interface Hamiltonian with a unique gapped ground state by adding local interaction terms $H_{\text{int}}$ around the interface,
\begin{equation}
    H_{\text{int}}=Y_{L-2}^{-1}X_{L-1}Z_{L}+Z_{\ell}^{-1}X_{\ell+1}Y_{\ell+2}+h.c..
\end{equation}
These interaction terms preserve the $\mathbb{Z}_N^{\rm e}\times\mathbb{Z}_N^{\rm o}$ symmetry but do not preserve the non-invertible symmetry. Therefore, it is clear that the $N^2$-fold degeneracy arises from the difference as the non-invertible SPT phases. 
Similar arguments can be applied to $\ket{\text{even}_m}$ as well.


\section{Conclusion and discussion}
In this paper, we studied an $N$-ality symmetry obtained by gauging a non-anomalous subgroup of $\mathbb{Z}_N\times\mathbb{Z}_N\times\mathbb{Z}_N$ symmetry with a type III anomaly. We construct the $N$-ality defects and derive the general fusion rules of the $N$-ality symmetry. Furthermore, as an application, we classify the SPT phases with the $N$-ality symmetry and explicitly construct $(2N-1)$ lattice Hamiltonians.
The relation of symmetries is given by the following figure \refeq{figure2}, where $G=(\mathbb{Z}_N^{\rmA} \times \mathbb{Z}_N^{\rmB}) \rtimes_{\rho} \mathbb{Z}_N^{\rmC}$ and $\mathcal{C}=\Rep(G)$.
\begin{figure}[H]
    \centering 
    \tikzset{none/.style={}} 
\newcommand{\bZ}{\mathbb{Z}} 

\begin{tikzpicture}
    \node [style=none] (0) at (-4, 0) {$\underbrace{\bZ_N^{\rmA}\times \bZ_N^{\rmB}\times \bZ_N^{\rmC}}_{\text{Type III anomaly}}$};
    \node [style=none] (A) at (-4.2, 0.8) {};
    \node [style=none] (A') at (-4.1, 0.6) {};
    \node [style=none] (3) at (-3.9, 2) {gauge $\bZ_N^{\rmA}$};
    \node [style=none] (4) at (-1.4, 2) {gauge $\bZ_N^{\rmA}$};

    \node [style=none] (C) at (-2.6, 0.3) {};
    \node [style=none] (C') at (-2.6, 0.1) {};
    
    \node [style=none] (1) at (0, 4) {$(\bZ_N^{\rmA} \times \bZ_N^{\rmB}) \rtimes_{\rho} \bZ_N^{\rmC}$};
    \node [style=none] (B) at (-0.9, 3.5) {};
    \node [style=none] (B') at (-0.8, 3.3) {};
    \node [style=none] (E) at (0.9, 3.5) {};
    \node [style=none] (E') at (0.8, 3.3) {};
    
    \node [style=none] (D) at (2.4, 0.3) {};
    \node [style=none] (D') at (2.4, 0.1) {};
    \node [style=none] (F) at (4.2, 1) {};
    \node [style=none] (F') at (4.1, 0.8) {};
    \node [style=none] (5) at (0, 0.6) {gauge $\bZ_N^{\rmB}\times\bZ_N^{\rmC}$};
    \node [style=none] (6) at (0, -0.2) {gauge $\bZ_N^{\rmB}\times\bZ_N^{\rmC}$};

    \node [style=none] (5) at (1.5, 2) {gauge $G$};
    \node [style=none] (6) at (3.9, 2) {gauge $\mathcal{C}$};

    \node [style=none] (2) at (4.8, 0.2) {$\mathrm{Rep} \left((\bZ_N^{\rmA} \times \bZ_N^{\rmB}) \rtimes_{\rho} \bZ_N^{\rmC}\right)$};

    \draw [stealth-] (B.center) to (A.center);
    \draw [stealth-] (A'.center) to (B'.center);
    \draw [stealth-] (C.center) to (D.center);
    \draw [stealth-] (D'.center) to (C'.center);
    \draw [stealth-] (E.center) to (F.center);
    \draw [stealth-] (F'.center) to (E'.center);
\end{tikzpicture}
    \caption{The relation of symmetries.}
    \label{figure2}
\end{figure}
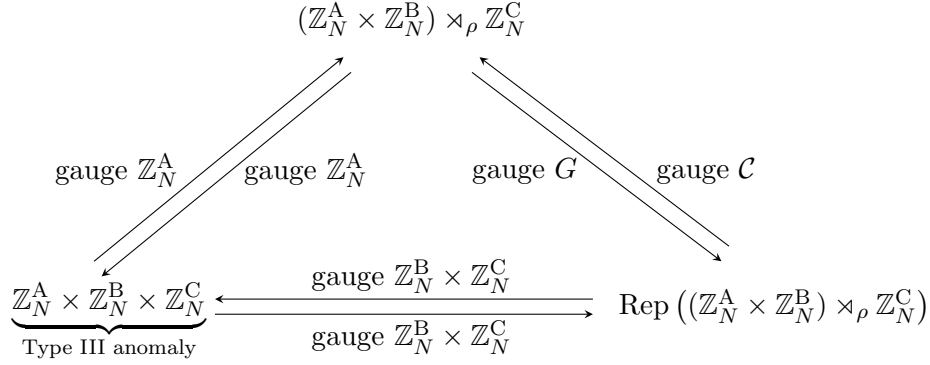

Finally, we list some interesting future directions.
\begin{itemize}
    \item What kind of $N$-ality category can be obtained from other types of mixed anomaly of $\mathbb{Z}_N\times\mathbb{Z}_N\times\mathbb{Z}_N$? For instance, can we obtain an anomalous $N$-ality category from type I+III anomaly? In the case $N=2$, it was discussed in \cite{Kaidi:2023maf}.

    \item How can we detect the data of F-symbols from the expression of the non-invertible defects we constructed? In \cite{Seifnashri:2025fgd}, for example, they develop the way to extract the data of F-symbols on the lattice.

    \item Generalization to higher dimensions. See \cite{Choi:2022jqy,Cordova:2022ieu,Hayashi:2022fkw} for $N$-ality symmetries in $(3+1)$d.

    \item We have completely classified SPT phases of the $N$-ality symmetry. However, we have not yet constructed all corresponding lattice Hamiltonians. Furthermore, other tensor product Hilbert space construction of $\Rep(D_8)$ SPT phases was studied in \cite{Warman:2024lir}. It’s a very interesting question whether this approach can be extended to the $N$-ality symmetries we discuss.

    \item It would be interesting to study the interface between different non-invertible SPT phases. We can distinguish between different non-invertible SPT phases by studying the action of non-invertible operators on the interface.

    \item What is the lattice realization of the $N$-ality defects? It would be interesting to find the matrix product operator (MPO) representation for the $N$-ality defects. The topological manipulation \textbf{S} can be realize as the MPO representation on the lattice \cite{Seiberg:2024gek,Zhang:2024nbt}.



\end{itemize}

\section*{Acknowledgements}
We thank Takamasa Ando, Hiromi Ebisu, Yuya Tanizaki, and Tatsuya Wada for useful discussions on this work. We thank Takamasa Ando, Hiromi Ebisu, Masatoshi Sato ,Yuya Tanizaki for useful comments on a draft.


\begin{appendices}
\section{Representation theory} \label{appendix A}
In this appendix, we demonstrate that the $N$-ality category discussed in this paper can be regarded as a $\Rep$-category of the following group $G$:
\begin{equation}
    G = \left( \mathbb{Z}_N^{\rmA}\times\mathbb{Z}_N^{\rmB} \right) \rtimes_{\rho} \mathbb{Z}_N^{\rmC},
\end{equation}
where $\rho(c)(a,b) = (a-cb, b)$. This group is regarded as the symmetry of the theory after gauging $\mathbb{Z}_N^{\rmA}$ of the symmetry $\mathbb{Z}_N^{\rmA}\times\mathbb{Z}_N^{\rmB}\times\mathbb{Z}_N^{\rmC}$ with a type III anomaly. In the case $N=2$, this group $G$ is isomorphic to $D_8$. Gauging the subgroup $\mathbb{Z}_N^{\rmB}\times\mathbb{Z}_N^{\rmC}\subset\mathbb{Z}_N^{\rmA}\times\mathbb{Z}_N^{\rmB}\times\mathbb{Z}_N^{\rmC}$ with a type III anomaly is equivalent to gauging group $G$, and thus the $N$-ality category we discuss is equivalent to $\Rep(G)$. Symmetry defects of the $N$-ality category correspond to the irreducible representations of $G$.

The irreducible representation of $G$ is labeled by a devisor $d$ of $N$, coprime $r$ to $d$, and $s,t\in\mathbb{Z}_{N/d}$ and we denote it by $\bm{d}^r_{s,t}$. The dimension of $\bm{d}^r_{s,t}$ is $d$ and the explicit form of it is given by
\begin{equation}
    \begin{cases}
        \begin{split}
            \bm{d}^r_{s,t}(a,0,0) &= e^{2\pi i\frac{ra}{d}}\,1_d \\
            \bm{d}^r_{s,t}(0,b,0) &= e^{2\pi i\frac{sb}{N}}\,S_d^b \\
            \bm{d}^r_{s,t}(0,0,c) &= e^{2\pi i\frac{tc}{N}}\,C_d^{rc},
        \end{split}
    \end{cases}
\end{equation}
where $1_d,S_d, C_d$ are $d$-dimensional unit matrix, shift matrix, and clock matrix, respectively. The total dimension of $\Rep(G)$ is
\begin{equation}
    \sum_{d|N}d^2\times\left(\frac{N}{d}\right)^2\times \varphi(d) = N^3,
\end{equation}
equivalent to that of $G$\footnote{$\varphi(d)$ is the Euler's function, which counts the number of positive integers coprime to $d$.}.

Next, we compute the fusion rules of $\Rep(G)$. Most of the part of the fusion rules are determined by the grading as the $N$-ality category. The grading of $\bm{d}^r_{s,t}$ is given by $\frac{Nr}{d}$. Therefore, the grading of $\bm{d}^r_{s,t}\otimes \tilde{\bm{d}}^{\tilde{r}}_{\tilde{s},\tilde{t}}$ is
\begin{equation}
    N\left(\frac{r}{d}+\frac{\tilde{r}}{\tilde{d}}\right) = \frac{NR}{D},
\end{equation}
where $D,R\in\mathbb{N}$ are coprime. It implies that the fusion is given by
\begin{equation}
    \bm{d}^r_{s,t}\otimes \tilde{\bm{d}}^{\tilde{r}}_{\tilde{s},\tilde{t}} = \bigoplus_{S,T} \bm{D}^R_{S,T}.
\end{equation}
However, the explicit values of $S,T$ and the range of the direct sum are not determined by the grading argument.

To determine this, we consider characters of the irreducible representations. The characters of $\bm{d}^r_{s,t}$ are given by
\begin{equation}\label{eq:caharacter of irrep}
    \chi_{\bm{d}^r_{s,t}}(a,b,c) = 
    \begin{cases}
        d \times \exp\left[2\pi i\left(\frac{ra}{d}+\frac{sb+tc}{N}\right)\right] & b,c \equiv 0\  \mathrm{mod}\ d \\
        0 & \text{otherwise}.
    \end{cases}
\end{equation}
The characters satisfy the following orthogonal relations:
\begin{equation}
    \langle\chi_1,\chi_2\rangle \coloneqq \frac{1}{|G|}\sum_{g\in G}\chi_1(g)\chi_2(g)^* = 
    \begin{cases}
        1 & \chi_1 = \chi_2 \\
        0 & \chi_1\neq \chi_2,
    \end{cases}
\end{equation}
where $\chi_1,\chi_2$ are irreducible representations of $G$. Therefore, we can compute the fusion coefficient by calculating
\begin{equation}
    \left\langle \chi_{\bm{d}^r_{s,t}\otimes \tilde{\bm{d}}^{\tilde{r}}_{\tilde{s},\tilde{t}}}, \chi_{\bm{D}^R_{S,T}} \right\rangle.
\end{equation}
Substituting \eqref{eq:caharacter of irrep} for this, the summand is nonzero only when $b,c$ are multiples of $l=\mathrm{lcm}(d,\tilde{d},D)$ and we get
\begin{equation}
    \begin{split}
        &\left\langle \chi_{\bm{d}^r_{s,t}\otimes \tilde{\bm{d}}^{\tilde{r}}_{\tilde{s},\tilde{t}}}, \chi_{\bm{D}^R_{S,T}} \right\rangle \\
        &= \frac{1}{N^3}\sum_{a=0}^{N-1}\sum_{l|b}\sum_{l|c} d\tilde{d}D\,\exp\left[2\pi i\left\{\left(\frac{r}{d}+\frac{\tilde{r}}{\tilde{d}}-\frac{R}{D}\right)a + \frac{s+\tilde{s}-S}{N}b + \frac{t+\tilde{t}-T}{N}c\right\}\right] \\
        &= \frac{d\tilde{d}D}{N^3}\sum_{a=0}^{N-1}\sum_{b = 0}^{N/l-1}\sum_{c=0}^{N/l-1}\exp\left[\frac{2\pi il}{N}\left\{(s+\tilde{s}-S)b + (t+\tilde{t}-T)c\right\}\right] \\
        &= \frac{d\tilde{d}D}{{l^2}}\delta^{(N/l)}_{s+\tilde{s}-S,0}\delta^{(N/l)}_{t+\tilde{t}-T,0},
    \end{split}
\end{equation}
where
\begin{equation}
    \delta^{(k)}_{a,b} = 
    \begin{cases}
        1 & a \equiv b\ \mathrm{mod}\ k \\
        0 & a \not\equiv b \ \mathrm{mod}\ k.
    \end{cases}
\end{equation}

Therefore, the fusion rules of $\Rep(G)$ are
\begin{equation}\label{eq:fusion rules of Rep(G)}
    \bm{d}^r_{s,t}\otimes \tilde{\bm{d}}^{\tilde{r}}_{\tilde{s},\tilde{t}} = \frac{d\tilde{d}D}{l^2}\bigoplus_{i,j=0}^{l/D-1} \bm{D}^R_{s+\tilde{s}+\frac{N}{l}i,t+\tilde{t}+\frac{N}{l}j}\ .
\end{equation}


\section{Topological manipulations on the lattice} \label{appendix B}
In this appendix, we briefly explain the topological manipulations on the lattice, used in this paper. 
\subsection{Gauging (\textbf{S})}
We review the (untwisted) gauging of $\mathbb{Z}_N$ symmetry in quantum spin system (see, e.g., 
\cite{Moradi:2023dan,Seifnashri:2023dpa}). For simplicity, we consider $\mathbb{Z}_N$ clock model as an example, but the procedure of gauging does not depend on the detail of symmetric Hamiltonians. The Hamiltonian of $\mathbb{Z}_N$ clock model is given by
\begin{equation}
    H_{\text{clock}} = -\sum_j(X_j+JZ_jZ_{j+1}^{\dagger}) + h.c.,
\end{equation}
where $J>0$ and $X,Z$ are the shift operator, clock operator, respectively which satisfy the following relations:
\begin{equation}
    X_j^N=Z_j^N=I,\quad Z_iX_j=\omega^{\delta_{ij}}X_jZ_i,\quad \omega=e^{\frac{2\pi i}{N}}, \label{clock_alg}
\end{equation}
where $I$ is the $N\times N$ unit matrix. 
This Hamiltonian has a $\mathbb{Z}_N$ global symmetry\footnote{Note that the $\mathbb{Z}_N$ global symmetry generated by $\eta$ is anomaly-free since it is an onsite zero form symmetry and one can construct a unique gapped phase.} generated by
\begin{equation}
    \eta=\prod_jX_j.
\end{equation}
To gauge this on the lattice, we introduce dual operators $\widetilde{X}_{j+\frac{1}{2}},\widetilde{Z}_{j+\frac{1}{2}}$ which satisfy 
\eqref{clock_alg} on the dual lattice. Next we define the Gauss law operator as 
\begin{equation}
    G_j=\widetilde{Z}_{j-\frac{1}{2}}X_j\widetilde{Z}_{j+\frac{1}{2}}^\dagger.
\end{equation}
The Gauss law operator generates a local $\mathbb{Z}_N$ symmetry in the sense that it satisfies $G_j^N=1$ and $\prod_jG_j=\eta$. The gauged Hamiltonian is defined by
\begin{equation}
    H_{\text{clock}}^{g}=-\sum_j(X_j+JZ_j\widetilde{X}_{j+\frac{1}{2}}^\dagger Z_{j+1}^{\dagger})+h.c..
\end{equation}
and it commutes with the Gauss law operator. Finally, we must impose the Gauss law constraint that the physical states $\ket{\psi}$ should be invariant under the action of $G_j$, namely $G_j\ket{\psi}=\ket{\psi}, \forall j$. Furthermore, to simplify the Gauss law operator, we implement the unitary transformation
\begin{equation}
    V_0=\prod_jCZ_{j,j-\frac{1}{2}}^\dagger CZ_{j,j+\frac{1}{2}},
\end{equation}
where $CZ$ is the controlled-$Z$ gate for $\mathbb{Z}_N$, defined by
\begin{equation}
    CZ_{i,j}=\frac{1}{N}\sum_{a=0}^{N-1}\sum_{b=0}^{N-1}\omega^{-ab}Z_i^aZ_j^b.
\end{equation}
The action of $V_0$ on the local operators is 
\begin{equation}
\begin{aligned}  
    X_j&\rightarrow V_0X_jV_0^\dagger=\widetilde{Z}_{j-\frac{1}{2}}^\dagger X_j\widetilde{Z}_{j+\frac{1}{2}},\quad&Z_j&\rightarrow Z_j,\\
    \widetilde{X}_{j+\frac{1}{2}}&\rightarrow V_0\widetilde{X}_{j+\frac{1}{2}}V_0^\dagger =Z_j\widetilde{X}_{j+\frac{1}{2}}Z_{j+1}^\dagger,\quad &\widetilde{Z}_{j+\frac{1}{2}}&\rightarrow\widetilde{Z}_{j+\frac{1}{2}}.
\end{aligned}
\end{equation}
Under this unitary transformation, the Gauss law operator is transformed as
\begin{equation}
   G_j\rightarrow  V_0 G_jV_0^\dagger=X_j,\quad \forall j.
\end{equation}
Thus, the gauged Hamiltonian is equivalent to 
\begin{equation}
    \widetilde{H}_{\text{clock}}^{g}=V_0 H_{\text{clock}}^{g}V_0^\dagger=-\sum_j(\widetilde{Z}_{j-\frac{1}{2}}^{\dagger}\widetilde{Z}_{j+\frac{1}{2}}+J\widetilde{X}_{j+\frac{1}{2}}^{\dagger})+h.c..
\end{equation}
By shifting the lattice site $j+\frac{1}{2}\rightarrow j$ and identifying the dual operators with original ones $\widetilde{X}=X,\widetilde{Z}=Z$, we obtain the gauged Hamiltonian on the original lattice
\begin{equation}
    \widetilde{H}_{\text{clock}}^{g}=-\sum_j(Z_{j-1}^{\dagger}Z_{j}+JX_{j}^{\dagger})+h.c..
\end{equation}
Therefore, the gauging of the $\mathbb{Z}_N$ symmetry is realized as the following transformation\footnote{This transformation cannot be implemented by a unitary operator.}
\begin{equation}
    X_j\rightsquigarrow Z_{j-1}^{\dagger}Z_{j},\quad Z_{j}^\dagger Z_{j+1}\rightsquigarrow X_{j}. \label{Z_N gauging}
\end{equation}
This transformation is well-known as the generalized Kramers-Wannier duality. The $\mathbb{Z}_N$ clock model at $J=1$ is invariant under the generalized Kramers-Wannier duality transformation \eqref{Z_N gauging}. Therefore, this transformation is a symmetry at $J=1$, described by the $\mathbb{Z}_N$ Tambara-Yamagami fusion category \cite{TAMBARA1998692}.

Let us apply the above discussion to $\mathbb{Z}_N^{\rm e}\times \mathbb{Z}_N^{\rm o}$ symmetry generated by
\begin{equation}
    \eta_e=\prod_{j:\text{even}}X_j,\quad \eta_o=\prod_{j:\text{odd}}X_j.
\end{equation}
We assume that the lattice size $L$ is even. The gauging of $\mathbb{Z}_N^{\rm e}\times \mathbb{Z}_N^{\rm o}$ symmetry is equivalent to gauging two $\mathbb{Z}_N$ symmetries independently. The gauging of $\mathbb{Z}_N^{\rm e}$ and $\mathbb{Z}_N^{\rm o}$ symmetries is realized as following transformations, respectively
\begin{equation}
\begin{aligned}
    D_e&:X_{2n}\rightsquigarrow Z_{2n-2}^\dagger Z_{2n},\quad Z_{2n-2}^\dagger Z_{2n}\rightsquigarrow X_{2n-2} \\
    D_o&:X_{2n+1}\rightsquigarrow Z_{2n-1}^\dagger Z_{2n+1},\quad Z_{2n-1}^\dagger Z_{2n+1}\rightsquigarrow X_{2n-1}.    \label{Z_N^2 gauging}
\end{aligned}
\end{equation}
Then, we can find the non-invertible operator 
\begin{equation}
    D=TD_eD_o,
\end{equation}
 where $T$ is a lattice translation operator. This non-invertible operator acts on local operators as
\begin{equation} \label{gauging}
    D:X_j\rightsquigarrow Z_{j-1}^\dagger Z_{j+1},\quad Z_{j-1}^\dagger Z_{j+1}\rightsquigarrow X_j ,
\end{equation}
and satisfies the following fusion rules
\begin{equation} \label{Z_N^2_fusion}
        D\times D=\sum_{m=0}^{N-1}\sum_{n=0}^{N-1}\eta_e^m\eta_o^n,\quad D\times \eta_e=\eta_e\times D=D\times \eta_o=\eta_o\times D=D.
\end{equation}
These fusion rules are described by the $\mathbb{Z}_N\times\mathbb{Z}_N$ Tambara-Yamagami fusion category \cite{TAMBARA1998692}. Thus, the gauging of $\mathbb{Z}_N^{\rm e}\times\mathbb{Z}_N^{\rm o}$ symmetry (\textbf{S}) is realized by the non-invertible operator $D$.


\subsection{Stacking SPT (\textbf{T})}
We review the stacking of an SPT (\textbf{T}) in $(1+1)$-dimensional lattice. Here, we consider
$\mathbb{Z}_N\times\mathbb{Z}_N$ SPT phases classified by the second group cohomology $H^2(\mathbb{Z}_N\times\mathbb{Z}_N;U(1))\simeq\mathbb{Z}_N=\{0,1,\cdots,N-1\}$ \cite{Chen:2011pg}. The Hamiltonian realizing the level-$k$ SPT phase \cite{Chen:2011pg,Tsui:2017ryj} is given by
\begin{equation} \label{k-SPT}
    H_{\text{SPT}_k}=-\sum_{n=1}^{L/2}(Z_{2n-1}^kX_{2n}Z_{2n+1}^{-k}+Z_{2n}^{-k}X_{2n+1}Z_{2n+2}^k)+h.c..
\end{equation}
This system has a $\mathbb{Z}_N\times\mathbb{Z}_N$ symmetry generated by
\begin{equation}
    \eta_e=\prod_{j:\text{even}}X_j,\quad \eta_o=\prod_{j:\text{odd}}X_j.
\end{equation}
Here, we assume that the system is defined on a periodic chain with an even number of sites $L$.
The Hamiltonian has a unique gapped ground state, denoted by $\ket{\text{SPT}_k}$ stabilized by the following $L$ generators
\begin{equation}
    Z_{2n-1}^kX_{2n}Z_{2n+1}^{-k}=1,\quad Z_{2n}^{-k}X_{2n+1}Z_{2n+2}^k=1,\quad \forall n.
\end{equation}
The SPT Hamiltonian can be obtained from the trivial Hamiltonian $(k=0)$ 
\begin{equation}
    H_{\text{trivial}}:=H_{\text{SPT}_0}=-\sum_{j=1}^L(X_j+X_j^\dagger)
\end{equation}
by the unitary transformation, called SPT entangler. The SPT entangler $V$ is given by 
\begin{equation}
    V=\prod_{n=1}^{L/2}CZ_{2n-1,2n}CZ_{2n,2n+1}^\dagger.
\end{equation}
One can check the following relations
\begin{equation}
    H_{\text{SPT}_k}=V^kH_{\text{trivial}}V^{-k},\quad V^N=1,
\end{equation} 
by using the action of $V$ on the local operators as
\begin{equation} \label{stacking SPT}
\begin{aligned}
    X_{2n}&\rightarrow VX_{2n}V^\dagger=Z_{2n-1}X_{2n}Z_{2n+1}^\dagger,\quad Z_{2n}\rightarrow Z_{2n},\\
    X_{2n+1}&\rightarrow VX_{2n+1}V^\dagger=Z_{2n}^\dagger X_{2n+1}Z_{2n+2},\quad Z_{2n+1}\rightarrow Z_{2n+1}.
\end{aligned}
\end{equation}
Thus, the stacking of an SPT (\textbf{T}) is realized by the unitary operator $V$.


\subsection{Others}
We explain the twisted gauging of $\mathbb{Z}_N^{\rm e}\times\mathbb{Z}_N^{\rm o}$ symmetry (\textbf{TST}) in quantum spin model. We also discuss the topological manipulation \textbf{TSTST$^{-1}$}.

The twisted gauging of $\mathbb{Z}_N^{\rm e}\times\mathbb{Z}_N^{\rm o}$ symmetry (\textbf{TST}) is realized by the combination of \eqref{gauging}, \eqref{stacking SPT} and implements the following transformation,
\begin{equation} \label{Z_N^2 twisted_gauging}
\begin{split}
        X_{2n}&\rightsquigarrow Z_{2n-1}^{-2}X_{2n}^{-1}Z_{2n+1}^2,\quad Z_{2n}^\dagger Z_{2n+2}\rightsquigarrow Z_{2n}^{-1}X_{2n+1}Z_{2n+2} \\
        X_{2n+1}&\rightsquigarrow Z_{2n}^{-2}X_{2n+1}Z_{2n+2}^2,\quad Z_{2n-1}^\dagger Z_{2n+1}\rightsquigarrow Z_{2n-1}X_{2n}Z_{2n+1}^{-1}.
\end{split}
\end{equation}
This transformation is the non-local mapping among different gapped phase with $\mathbb{Z}_N^{\rm e}\times\mathbb{Z}_N^{\rm o}$ symmetry. For example, the $\mathbb{Z}_N^{\rm e}\times\mathbb{Z}_N^{\rm o}$ SSB phase is mapped to the level-1 SPT phase as follows
\begin{equation}  
\begin{aligned}
\begin{tikzpicture}
        \centering
        \node[] (1) at (0,0) {$ H_{\text{SSB}}=-\sum_{n=1}^{L/2}(Z_{2n-1}Z_{2n+1}^{\dagger}+Z_{2n}Z_{2n+2}^{\dagger})+h.c.$,}; 
        \node[] (2) at (0,-1.5){ $ H_{\text{SPT}_1}=-\sum_{n=1}^{L/2}(Z_{2n-1}X_{2n}Z_{2n+1}^{-1}+Z_{2n}^{-1}X_{2n+1}Z_{2n+2})+h.c.$.};
        \draw[->] (1) -- node[scale=.8,anchor=west]{\ Twisted gauging $\textbf{TST}$} (2);
    \end{tikzpicture}
    \end{aligned}
\end{equation}
In the case of $N=2$, this transformation corresponds to Kennedy-Tasaki transformation \cite{Li:2023ani,Kennedy:1992ifl,Kennedy:1992tke, M.Oshikawa_1992}. 

Finally, the topological manipulation $\textbf{TSTST}^{-1}$ is realized as the following transformation
\begin{equation}
\begin{aligned}
    X_{2n}&\rightsquigarrow Z_{2n-1}^{\dagger}Z_{2n+1},\quad \quad \quad \ \ \ \ \ \ X_{2n+1}\rightsquigarrow Z_{2n}Z_{2n+2}^{\dagger},\\
    Z_{2n-1}Z_{2n+1}^\dagger&\rightsquigarrow Z_{2n-1}^2 X_{2n} Z_{2n+1}^{-2},\quad Z_{2n}Z_{2n+2}^\dagger\rightsquigarrow Z_{2n}^2 X_{2n+1}^\dagger Z_{2n+2}^{-2}
\end{aligned}
\end{equation}
and satisfies $(\textbf{TSTST}^{-1})^N=1$, where we have used $\textbf{S}^2=1$ and $\textbf{T}^N=1$.


\section{Type III anomaly and Non-Abelian group} \label{appendix C}
In this appendix, we discuss the lattice descriptions of a type III anomaly, and show that a non-Abelian group can be obtained by gauging a non-anomalous subgroup of $\mathbb{Z}_N^{\rm V}\times\mathbb{Z}_N^{\rm e}\times\mathbb{Z}_N^{\rm o}$.


\subsection{Projective algebras from type III anomaly}
We have discussed the $\mathbb{Z}_N^{\rm V}\times\mathbb{Z}_N^{\rm e}\times\mathbb{Z}_N^{\rm o}$ symmetry generated by \eqref{sym_op}. Here, we show that it has a type III anomaly. We consider a symmetric Hamiltonian given by 
\begin{equation}
    H=-\sum_{n=1}^{L/2}\sum_{k=0}^{N-1}\qty(Z_{2n-1}^kX_{2n}Z_{2n+1}^{-k}+Z_{2n}^{-k} X_{2n+1}Z_{2n+2}^k)+h.c..
\end{equation}
Note that the anomalies are only determined by symmetry operators and do not depend on the choice of symmetric Hamiltonians. One method for detecting the type III anomaly is the projective algebra of two $\mathbb{Z}_N$ symmetry operators in the Hamiltonian twisted by the other $\mathbb{Z}_N$ symmetry. See \cite{Seifnashri:2023dpa, Else:2014vma, Kawagoe:2021gqi} for general methods for detecting the anomalies on the $1$d lattice.  We consider the Hamiltonian twisted by an $\eta_e$ defect at site $1$, 
\begin{align*}
    H_{{\eta}_e}=&-(X_{L}+Z_{L-1}X_{L}Z_{1}^{-1}+\cdots+Z_{L-1}^{N-1}X_{L}Z_{1}^{-(N-1)})\\
                &-(X_{1}+\omega Z_{L}^{-1}X_{1}Z_{2}+\cdots+\omega^{N-1}Z_{L}^{-(N-1)}X_{1}Z_{2}^{N-1})\\
                &-(X_{2}+Z_{1}X_{2}Z_{3}^{-1}+\cdots+Z_{1}^{N-1}X_{2}Z_{3}^{-(N-1)})\\
                &-\cdots.
\end{align*}
Upon inserting an $\eta_e$ defect, the Hamiltonian no longer commutes with $V$. However, it commutes with $\widetilde{V}=Z_1V$,
where $\widetilde{V}$ is obtained by modifying $V$ around the defect.
Then one can detect the projective algebra between $\eta_o$ and $\widetilde{V}$,
\begin{equation}
    \Tilde{V}\eta_o=\omega\eta_o\Tilde{V}
\end{equation}
and it characterizes the type III anomaly on the lattice.


\subsection{Non-Abelian group from gauging}
In this subsection, we show that a non-Abelian group can be obtained by gauging the $\mathbb{Z}_N^{\rm V}$ symmetry of $\mathbb{Z}_N^{\rm V}\times\mathbb{Z}_N^{\rm e}\times\mathbb{Z}_N^{\rm o}$.
To implement this gauging, we introduce dual operators $\widetilde{X}_j, \widetilde{Z}_j$ on sites, and then we define the Gauss law operator as
\begin{equation}
    G_{2n}=\widetilde{X}_{2n-1}\widetilde{X}_{2n}CZ_{2n-1,2n}CZ^\dagger_{2n,2n+1}\widetilde{X}_{2n+1}^\dagger\widetilde{X}_{2n+2}^\dagger.
\end{equation}
The Gauss law operator generates a local $\mathbb{Z}_N^{\rm V}$ symmetry in the sense that it satisfies $G_{2n}^N=1$ and $\prod_{n=1}^{L/2}G_{2n}=V$. By imposing a Gauss law constraint, we obtain a non-Abelian group symmetry generated by
\begin{align}
    \hat{V}=\prod_{j}\widetilde{Z}_j, \quad
    \hat{\eta}_e=\prod_{j:\text{even}}X_j\prod_{j:\text{odd}}CZ_{j,\tilde{j}},\quad
    \hat{\eta}_o=\prod_{j:\text{odd}}X_j\prod_{j:\text{even}}CZ^\dagger_{j,\tilde{j}},
\end{align}
where $CZ_{j,\tilde{j}}=\frac{1}{N}\sum_{a,b=0}^{N-1}\omega^{-ab}Z_j\widetilde{Z}_j$ and $\hat{V}$ is a dual $\mathbb{Z}_N^{\rm V}$ symmetry generator. One can check that these operators commute with the Gauss law operator, and satisfy the following algebras,
\begin{align}
    \hat{\eta}_e^N=\hat{\eta}_o^N=\hat{V}^N=1,\quad \hat{\eta}_e\hat{V}=\hat{V}\hat{\eta}_e,\quad \hat{\eta}_o\hat{V}=\hat{V}\hat{\eta}_o,\quad \hat{\eta}_o^{-1}\hat{\eta}_e\hat{\eta}_o=\hat{V}\hat{\eta}_e. 
\end{align}
Thus, this non-Abelian group $G$ is represented as 
\begin{equation} \label{non-abelian}
    G=\langle a,b,c\ |\ a^N=b^N=c^N=1,\ ab=ba,\ ca=ac,\ bc=acb\ \rangle.
\end{equation}
Here, we identify $\hat{V}$, $\hat{\eta}_e$ and $\hat{\eta}_o$ with $a,b$ and $c$, respectively. This non-Abelian group can be expressed in the form of a semi-direct product $(\mathbb{Z}_N^{\rmA}\times\mathbb{Z}_N^{\rmB})\rtimes_{\rho}\mathbb{Z}_N^{\rmC}$, where $\rho$ is a group homomorphism 
\begin{equation}
    \rho:\mathbb{Z}_N^{\rmC}\rightarrow \mathrm{Aut}(\mathbb{Z}_N^{\rmA}\times\mathbb{Z}_N^{\rmB}),\quad \rho(c)(a,b)=(a-cb,b).
\end{equation}

\section{Additional Computations} \label{appendix D}

\subsection{Gauging map of SPT phase}
In this appendix, we derive the expression \eqref{eq:S of level k SPT} of the partition function when ungauging the level-$k$ SPT phase of $\widehat{\mathbb{Z}}_N^{\rmB}\times\widehat{\mathbb{Z}}_N^{\rmC}$ symmetry, where
\begin{equation}
    N=dx,\quad k=dy,\quad \text{gcd}(N,k)=d,\quad \text{gcd}(x,y)=1.
\end{equation}
The partition function can be computed as follows,
\begin{equation}
    \begin{aligned}
        \frac{1}{|H^1(X;\mathbb{Z}_N)|}&\sum_{b,c\in H^1(X;\mathbb{Z}_N)}
        \exp[\frac{2\pi i}{N}\int kb\cup c + b\cup C + c\cup B]  \\
        &=\sum_{c\in H^1(X;\mathbb{Z}_N)} \delta^{(N)}(dyc+C)\exp[\frac{2\pi i}{N}\int c\cup B],
    \end{aligned}
\end{equation}
where 
\begin{equation}
    \delta^{(N)}(A)=
    \begin{cases}
     1     & \int A\equiv 0 \mod N \\
     0     & \int A\not\equiv 0 \mod N.
  \end{cases}
\end{equation}
This partition function vanishes if $\int C$ is nonzero modulo $d$. On the other hand, it does not vanish if $\int C$ is zero modulo $d$, and then one can take $C=d\widetilde{C}$, where $\widetilde{C}$ is a $\mathbb{Z}_x$ background gauge field. Then, we obtain
\begin{equation}
    \begin{aligned}
        \sum_{c\in H^1(X;\mathbb{Z}_N)} &\delta^{(x)}(yc+\widetilde{C})\exp[\frac{2\pi i}{N}\int c\cup B] \\
        =&\sum_{c'\in H^1(X;\mathbb{Z}_d)} \exp[\frac{2\pi i}{N}\int (-y^{-1}\widetilde{C} + xc')\cup B] \\
        =&|H^1(X;\mathbb{Z}_d)|\ \delta^{(N)}(xB)\exp[\frac{2\pi i}{x} y^{-1}\int \frac{B}{d}\cup \frac{C}{d}],
    \end{aligned}
\end{equation}
where $c'$ is a $\mathbb{Z}_d$ gauge field and $y^{-1}$ is the multiplicative inverse of $y$ modulo $x$. In the first equality, summing over $c$ enforces the constraint $c = -y^{-1} \widetilde{C} + xc'$. 
Thus, the partition function of gauged theory is given by
\begin{equation}
    |H^1(X;\mathbb{Z}_d)|\ \delta^{(N)}(xB)\delta^{(N)}(xC)\exp[\frac{2\pi i}{x} y^{-1}\int \frac{B}{d}\cup \frac{C}{d}].
\end{equation}

\subsection{Reduction of type III anomaly} 
In this appendix, we show that the anomalies $\exp[\frac{2\pi i}{N}\int A^3]$ and $\exp[\frac{2\pi i}{N}\int A^2\cup B]$ are trivial when $N$ is odd, and a non-trivial $\mathbb{Z}_2$ anomaly when $N$ is even.
To see this, we first apply a higher cup products formula to $A\cup A$,
\begin{equation}
        A\cup A=-A\cup A+\delta A\cup A+A\cup \delta A-\delta(A\cup_1 A),
\end{equation}
where $A$ is just a cochain. For the gauge field $A$, this formula leads to
\begin{equation}
    2 A\cup A = \delta(A\cup_1A),
\end{equation}
since the discrete gauge field $A$ is closed. Therefore, when $N$ is odd, $A^2B$ can be transformed into
\begin{equation}
    A\cup A\cup B=\frac{N+1}{2}\delta(A\cup_1 A \cup B). \label{3.4}
\end{equation}
In contrast, it cannot be transformed into \eqref{3.4} when $N$ is even.
Therefore, it follows that for odd $N$, $\mathrm{exp}\qty({\frac{2\pi i}{N}\int A^2 \cup B})$ is a trivial cocycle, and for even $N$, it is a non-trivial cocycle that gives $\mathbb{Z}_2$ anomaly. Similar arguments can be applied to $\exp[\frac{2\pi i}{N}\int A^3]$ as well.

\subsection{Stacking SPT phases associated with the unbroken symmetry}

As discussed in the main text, stacking SPT phases associated with the subgroups $\mathbb{Z}_N\times\mathbb{Z}_x^{\rmB}, \mathbb{Z}_N\times\mathbb{Z}_x^{\rmC}\subset K$ symmetry does not lead to new gapped phases. In what follows, we explain the reason for this in detail.

First, we state the general criterion for identifying gapped phases with a discrete group-like symmetry. Let us consider a symmetry group $G$ with an anomaly $\omega\in H^3(G;U(1))$. A gapped phase of this symmetry is characterized by a subgroup $K\subset G$ and $\psi_K\in H^2(K;U(1))$. According to Ref.~\cite{Sonia2017}, two gapped phases labeled by $(K,\psi_K)$ and $(L,\psi_L)$ are equivalent if and only if there exists an element $g\in G$ such that $K={}^gL\coloneqq\{{}^ga\mid a\in L\}$ and 
\begin{equation}\label{eq:condition for equiv}
    \psi_L^{-1}\psi_K^g\Omega_g|_L
\end{equation}
is trivial in $H^2(L;U(1))$, where ${}^ga\coloneqq gag^{-1}$ and 
\begin{equation}
    \begin{split}
        \psi_K^g(g_1,g_2) &\coloneqq \psi_K({}^gg_1.{}^gg_2) \\
        \Omega_g(g_1,g_2) &\coloneqq \frac{\omega({}^gg_1,{}^gg_2,g)\omega(g,g_1,g_2)}{\omega({}^gg_1,g,g_2)}.
    \end{split}
\end{equation}
Physically, this condition means that two phases $(K,\psi_K)$ and $(L,\psi_L)$ are related by inserting a topological line defect labeled by $g$.

In the case considered in the paper, the symmetry group is $G=\mathbb{Z}_N^{\rmA}\times\mathbb{Z}_N^{\rmB}\times\mathbb{Z}_N^{\rmC}$, and the anomaly is the type III cocycle:
\begin{equation}
    \omega((a_1,b_1,c_1),(a_2,b_2,c_2),(a_3,b_3,c_3))=e^{\frac{2\pi i}{N}a_1b_2c_3},
\end{equation}
where $(a_i,b_i,c_i)\in G$. In the main text, we focus on the subgroup $K=\langle(1,m,n), (0,d,0), (0,0,d)\rangle$ and $\psi_K$ that represents the level-$y^{-1}$ $\mathbb{Z}_x^{\rmB}\times\mathbb{Z}_x^{\rmC}$ SPT phase.

Let us determine the condition on $(L,\psi_L)$ such that it represents the same phase as $(K,\psi_K)$. First, since the conjugate action of the abelian group $G$ is trivial, we have $L=K$. Furthermore, as discussed in the main text, $\psi_L$ must include the level-$y^{-1}$ $\mathbb{Z}_x^{\rmB}\times\mathbb{Z}_x^{\rmC}$ SPT phase. However, $\psi_L$ may additionally include SPT phases associated with the symmetries $\mathbb{Z}_N\times\mathbb{Z}_x^{\rmB}$ and $\mathbb{Z}_N\times\mathbb{Z}_x^{\rmC}$. Let us denote their levels by $l_1, l_2\in\mathbb{Z}_x$, respectively. We now evaluate the value~\eqref{eq:condition for equiv} for $g=(a,b,c)\in G$ and
\begin{equation}
    g_i = (\alpha_i, m\alpha_i+d\beta_i, n\alpha_i+d\gamma_i)\in L,
\end{equation}
where $\alpha_i\in\mathbb{Z}_N$ and $\beta_i, \gamma_i\in\mathbb{Z}_x$. We obtain
\begin{equation}
    \begin{split}
        &(\psi_L^{-1}\psi_K^g\Omega_g)(g_1,g_2) = \\ 
        &\qquad \exp\left[\frac{2\pi i}{N}\{(mc+mna-nb)\alpha_1\alpha_2 + (dc+dl_1)\alpha_1\beta_2 + nda\beta_1\alpha_2 + (mda-db+dl_2)\alpha_1\gamma_2 \}\right].
    \end{split}
\end{equation}
Since this cocycle must be trivial in $H^2(L;U(1))$, we obtain the following conditions:
\begin{equation}
    \begin{split}
        na - c - l_1 &\equiv 0 \mod x \\
        ma - b + l_2 &\equiv 0 \mod x.
    \end{split}
\end{equation}
For any choice of $l_1, l_2\in\mathbb{Z}_x$, these equations can always be solved by an appropriate choice of $g=(a,b,c)\in G$. Therefore, $(L,\psi_L)$ represents the same phase as $(K,\psi_K)$.


\end{appendices}


\bibliography{ref.bib}

\providecommand{\href}[2]{#2}\begingroup\raggedright\begin{thebibliography}{10}

\bibitem{Gaiotto:2014kfa}
D.~Gaiotto, A.~Kapustin, N.~Seiberg and B.~Willett, \emph{{Generalized Global Symmetries}}, \href{https://doi.org/10.1007/JHEP02(2015)172}{\emph{JHEP} {\bfseries 02} (2015) 172} [\href{https://arxiv.org/abs/1412.5148}{{\ttfamily 1412.5148}}].

\bibitem{Bhardwaj:2017xup}
L.~Bhardwaj and Y.~Tachikawa, \emph{{On finite symmetries and their gauging in two dimensions}}, \href{https://doi.org/10.1007/JHEP03(2018)189}{\emph{JHEP} {\bfseries 03} (2018) 189} [\href{https://arxiv.org/abs/1704.02330}{{\ttfamily 1704.02330}}].

\bibitem{Chang:2018iay}
C.-M.~Chang, Y.-H.~Lin, S.-H.~Shao, Y.~Wang and X.~Yin, \emph{{Topological Defect Lines and Renormalization Group Flows in Two Dimensions}}, \href{https://doi.org/10.1007/JHEP01(2019)026}{\emph{JHEP} {\bfseries 01} (2019) 026} [\href{https://arxiv.org/abs/1802.04445}{{\ttfamily 1802.04445}}].

\bibitem{Shao:2023gho}
S.-H.~Shao, \emph{{What's Done Cannot Be Undone: TASI Lectures on Non-Invertible Symmetries}},  \href{https://arxiv.org/abs/2308.00747}{{\ttfamily 2308.00747}}.

\bibitem{Schafer-Nameki:2023jdn}
S.~Schafer-Nameki, \emph{{ICTP lectures on (non-)invertible generalized symmetries}}, \href{https://doi.org/10.1016/j.physrep.2024.01.007}{\emph{Phys. Rept.} {\bfseries 1063} (2024) 1} [\href{https://arxiv.org/abs/2305.18296}{{\ttfamily 2305.18296}}].

\bibitem{Kramers:1941kn}
H.A.~Kramers and G.H.~Wannier, \emph{{Statistics of the two-dimensional ferromagnet. Part 1.}}, \href{https://doi.org/10.1103/PhysRev.60.252}{\emph{Phys. Rev.} {\bfseries 60} (1941) 252}.

\bibitem{TAMBARA1998692}
D.~Tambara and S.~Yamagami, \emph{Tensor categories with fusion rules of self-duality for finite abelian groups}, \href{https://doi.org/https://doi.org/10.1006/jabr.1998.7558}{\emph{Journal of Algebra} {\bfseries 209} (1998) 692}.

\bibitem{Thorngren:2019iar}
R.~Thorngren and Y.~Wang, \emph{{Fusion category symmetry. Part I. Anomaly in-flow and gapped phases}}, \href{https://doi.org/10.1007/JHEP04(2024)132}{\emph{JHEP} {\bfseries 04} (2024) 132} [\href{https://arxiv.org/abs/1912.02817}{{\ttfamily 1912.02817}}].

\bibitem{Bhardwaj:2023fca}
L.~Bhardwaj, L.E.~Bottini, D.~Pajer and S.~Schafer-Nameki, \emph{{Categorical Landau Paradigm for Gapped Phases}}, \href{https://doi.org/10.1103/PhysRevLett.133.161601}{\emph{Phys. Rev. Lett.} {\bfseries 133} (2024) 161601} [\href{https://arxiv.org/abs/2310.03786}{{\ttfamily 2310.03786}}].

\bibitem{Bhardwaj:2023idu}
L.~Bhardwaj, L.E.~Bottini, D.~Pajer and S.~Sch\"afer-Nameki, \emph{{Gapped Phases with Non-Invertible Symmetries: (1+1)d}}, \href{https://doi.org/10.21468/SciPostPhys.18.1.032}{\emph{SciPost Phys.} {\bfseries 18} (2025) 032} [\href{https://arxiv.org/abs/2310.03784}{{\ttfamily 2310.03784}}].

\bibitem{Bhardwaj:2024qiv}
L.~Bhardwaj, D.~Pajer, S.~Schafer-Nameki, A.~Tiwari, A.~Warman and J.~Wu, \emph{{Gapped Phases in (2+1)d with Non-Invertible Symmetries: Part I}},  \href{https://arxiv.org/abs/2408.05266}{{\ttfamily 2408.05266}}.

\bibitem{Bhardwaj:2025piv}
L.~Bhardwaj, S.~Schafer-Nameki, A.~Tiwari and A.~Warman, \emph{{Gapped Phases in (2+1)d with Non-Invertible Symmetries: Part II}},  \href{https://arxiv.org/abs/2502.20440}{{\ttfamily 2502.20440}}.

\bibitem{Chen:2010zpc}
X.~Chen, Z.-C.~Gu and X.-G.~Wen, \emph{{Classification of gapped symmetric phases in one-dimensional spin systems}}, \href{https://doi.org/10.1103/PhysRevB.83.035107}{\emph{Phys. Rev. B} {\bfseries 83} (2011) 035107} [\href{https://arxiv.org/abs/1008.3745}{{\ttfamily 1008.3745}}].

\bibitem{Chen:2011pg}
X.~Chen, Z.-C.~Gu, Z.-X.~Liu and X.-G.~Wen, \emph{{Symmetry protected topological orders and the group cohomology of their symmetry group}}, \href{https://doi.org/10.1103/PhysRevB.87.155114}{\emph{Phys. Rev. B} {\bfseries 87} (2013) 155114} [\href{https://arxiv.org/abs/1106.4772}{{\ttfamily 1106.4772}}].

\bibitem{Chen:2010gda}
X.~Chen, Z.C.~Gu and X.G.~Wen, \emph{{Local unitary transformation, long-range quantum entanglement, wave function renormalization, and topological order}}, \href{https://doi.org/10.1103/PhysRevB.82.155138}{\emph{Phys. Rev. B} {\bfseries 82} (2010) 155138} [\href{https://arxiv.org/abs/1004.3835}{{\ttfamily 1004.3835}}].

\bibitem{Gu:2009dr}
Z.-C.~Gu and X.-G.~Wen, \emph{{Tensor-Entanglement-Filtering Renormalization Approach and Symmetry Protected Topological Order}}, \href{https://doi.org/10.1103/PhysRevB.80.155131}{\emph{Phys. Rev. B} {\bfseries 80} (2009) 155131} [\href{https://arxiv.org/abs/0903.1069}{{\ttfamily 0903.1069}}].

\bibitem{Levin:2012yb}
M.~Levin and Z.-C.~Gu, \emph{{Braiding statistics approach to symmetry-protected topological phases}}, \href{https://doi.org/10.1103/PhysRevB.86.115109}{\emph{Phys. Rev. B} {\bfseries 86} (2012) 115109} [\href{https://arxiv.org/abs/1202.3120}{{\ttfamily 1202.3120}}].

\bibitem{Pollmann:2009mhk}
F.~Pollmann, E.~Berg, A.M.~Turner and M.~Oshikawa, \emph{{Symmetry protection of topological phases in one-dimensional quantum spin systems}}, \href{https://doi.org/10.1103/PhysRevB.85.075125}{\emph{Phys. Rev. B} {\bfseries 85} (2012) 075125} [\href{https://arxiv.org/abs/0909.4059}{{\ttfamily 0909.4059}}].

\bibitem{Pollmann:2009ryx}
F.~Pollmann, A.M.~Turner, E.~Berg and M.~Oshikawa, \emph{{Entanglement spectrum of a topological phase in one dimension}}, \href{https://doi.org/10.1103/PhysRevB.81.064439}{\emph{Phys. Rev. B} {\bfseries 81} (2010) 064439} [\href{https://arxiv.org/abs/0910.1811}{{\ttfamily 0910.1811}}].

\bibitem{Inamura:2021wuo}
K.~Inamura, \emph{{Topological field theories and symmetry protected topological phases with fusion category symmetries}}, \href{https://doi.org/10.1007/JHEP05(2021)204}{\emph{JHEP} {\bfseries 05} (2021) 204} [\href{https://arxiv.org/abs/2103.15588}{{\ttfamily 2103.15588}}].

\bibitem{Seifnashri:2024dsd}
S.~Seifnashri and S.-H.~Shao, \emph{{Cluster State as a Noninvertible Symmetry-Protected Topological Phase}}, \href{https://doi.org/10.1103/PhysRevLett.133.116601}{\emph{Phys. Rev. Lett.} {\bfseries 133} (2024) 116601} [\href{https://arxiv.org/abs/2404.01369}{{\ttfamily 2404.01369}}].

\bibitem{Lu:2022ver}
D.-C.~Lu and Z.~Sun, \emph{{On triality defects in 2d CFT}}, \href{https://doi.org/10.1007/JHEP02(2023)173}{\emph{JHEP} {\bfseries 02} (2023) 173} [\href{https://arxiv.org/abs/2208.06077}{{\ttfamily 2208.06077}}].

\bibitem{Lu:2024lzf}
D.-C.~Lu, Z.~Sun and Z.~Zhang, \emph{{Exploring $G$-ality defects in 2-dim QFTs}},  \href{https://arxiv.org/abs/2406.12151}{{\ttfamily 2406.12151}}.

\bibitem{Ando:2024hun}
T.~Ando, \emph{{A journey on self-$G$-ality}},  \href{https://arxiv.org/abs/2405.15648}{{\ttfamily 2405.15648}}.

\bibitem{Lu:2024ytl}
D.-C.~Lu, Z.~Sun and Y.-Z.~You, \emph{{Realizing triality and $p$-ality by lattice twisted gauging in (1+1)d quantum spin systems}}, \href{https://doi.org/10.21468/SciPostPhys.17.5.136}{\emph{SciPost Phys.} {\bfseries 17} (2024) 136} [\href{https://arxiv.org/abs/2405.14939}{{\ttfamily 2405.14939}}].

\bibitem{Lu:2025gpt}
D.-C.~Lu, Z.~Sun and Z.~Zhang, \emph{{SymSETs and self-dualities under gauging non-invertible symmetries}},  \href{https://arxiv.org/abs/2501.07787}{{\ttfamily 2501.07787}}.

\bibitem{Etingof:2009yvg}
P.~Etingof, D.~Nikshych, V.~Ostrik and w.a.a.b.E.~Meir, \emph{{Fusion categories and homotopy theory}},  \href{https://arxiv.org/abs/0909.3140}{{\ttfamily 0909.3140}}.

\bibitem{Gelaki:2009blp}
S.~Gelaki, D.~Naidu and D.~Nikshych, \emph{{Centers of graded fusion categories}},  \href{https://arxiv.org/abs/0905.3117}{{\ttfamily 0905.3117}}.

\bibitem{Thorngren:2021yso}
R.~Thorngren and Y.~Wang, \emph{{Fusion category symmetry. Part II. Categoriosities at c = 1 and beyond}}, \href{https://doi.org/10.1007/JHEP07(2024)051}{\emph{JHEP} {\bfseries 07} (2024) 051} [\href{https://arxiv.org/abs/2106.12577}{{\ttfamily 2106.12577}}].

\bibitem{Choi:2022zal}
Y.~Choi, C.~Cordova, P.-S.~Hsin, H.T.~Lam and S.-H.~Shao, \emph{{Non-invertible Condensation, Duality, and Triality Defects in 3+1 Dimensions}}, \href{https://doi.org/10.1007/s00220-023-04727-4}{\emph{Commun. Math. Phys.} {\bfseries 402} (2023) 489} [\href{https://arxiv.org/abs/2204.09025}{{\ttfamily 2204.09025}}].

\bibitem{Choi:2022jqy}
Y.~Choi, H.T.~Lam and S.-H.~Shao, \emph{{Noninvertible Global Symmetries in the Standard Model}}, \href{https://doi.org/10.1103/PhysRevLett.129.161601}{\emph{Phys. Rev. Lett.} {\bfseries 129} (2022) 161601} [\href{https://arxiv.org/abs/2205.05086}{{\ttfamily 2205.05086}}].

\bibitem{Cordova:2022ieu}
C.~Cordova and K.~Ohmori, \emph{{Noninvertible Chiral Symmetry and Exponential Hierarchies}}, \href{https://doi.org/10.1103/PhysRevX.13.011034}{\emph{Phys. Rev. X} {\bfseries 13} (2023) 011034} [\href{https://arxiv.org/abs/2205.06243}{{\ttfamily 2205.06243}}].

\bibitem{Hayashi:2022fkw}
Y.~Hayashi and Y.~Tanizaki, \emph{{Non-invertible self-duality defects of Cardy-Rabinovici model and mixed gravitational anomaly}}, \href{https://doi.org/10.1007/JHEP08(2022)036}{\emph{JHEP} {\bfseries 08} (2022) 036} [\href{https://arxiv.org/abs/2204.07440}{{\ttfamily 2204.07440}}].

\bibitem{Kaidi:2023maf}
J.~Kaidi, E.~Nardoni, G.~Zafrir and Y.~Zheng, \emph{{Symmetry TFTs and anomalies of non-invertible symmetries}}, \href{https://doi.org/10.1007/JHEP10(2023)053}{\emph{JHEP} {\bfseries 10} (2023) 053} [\href{https://arxiv.org/abs/2301.07112}{{\ttfamily 2301.07112}}].

\bibitem{Tachikawa:2017gyf}
Y.~Tachikawa, \emph{{On gauging finite subgroups}}, \href{https://doi.org/10.21468/SciPostPhys.8.1.015}{\emph{SciPost Phys.} {\bfseries 8} (2020) 015} [\href{https://arxiv.org/abs/1712.09542}{{\ttfamily 1712.09542}}].

\bibitem{Nguyen:2021yld}
M.~Nguyen, Y.~Tanizaki and M.~\"Unsal, \emph{{Semi-Abelian gauge theories, non-invertible symmetries, and string tensions beyond $N$-ality}}, \href{https://doi.org/10.1007/JHEP03(2021)238}{\emph{JHEP} {\bfseries 03} (2021) 238} [\href{https://arxiv.org/abs/2101.02227}{{\ttfamily 2101.02227}}].

\bibitem{Heidenreich:2021xpr}
B.~Heidenreich, J.~McNamara, M.~Montero, M.~Reece, T.~Rudelius and I.~Valenzuela, \emph{{Non-invertible global symmetries and completeness of the spectrum}}, \href{https://doi.org/10.1007/JHEP09(2021)203}{\emph{JHEP} {\bfseries 09} (2021) 203} [\href{https://arxiv.org/abs/2104.07036}{{\ttfamily 2104.07036}}].

\bibitem{Kaidi:2021xfk}
J.~Kaidi, K.~Ohmori and Y.~Zheng, \emph{{Kramers-Wannier-like Duality Defects in (3+1)D Gauge Theories}}, \href{https://doi.org/10.1103/PhysRevLett.128.111601}{\emph{Phys. Rev. Lett.} {\bfseries 128} (2022) 111601} [\href{https://arxiv.org/abs/2111.01141}{{\ttfamily 2111.01141}}].

\bibitem{Hsin:2018vcg}
P.-S.~Hsin, H.T.~Lam and N.~Seiberg, \emph{{Comments on One-Form Global Symmetries and Their Gauging in 3d and 4d}}, \href{https://doi.org/10.21468/SciPostPhys.6.3.039}{\emph{SciPost Phys.} {\bfseries 6} (2019) 039} [\href{https://arxiv.org/abs/1812.04716}{{\ttfamily 1812.04716}}].

\bibitem{Ji:2019jhk}
W.~Ji and X.-G.~Wen, \emph{{Categorical symmetry and noninvertible anomaly in symmetry-breaking and topological phase transitions}}, \href{https://doi.org/10.1103/PhysRevResearch.2.033417}{\emph{Phys. Rev. Res.} {\bfseries 2} (2020) 033417} [\href{https://arxiv.org/abs/1912.13492}{{\ttfamily 1912.13492}}].

\bibitem{Gaiotto:2020iye}
D.~Gaiotto and J.~Kulp, \emph{{Orbifold groupoids}}, \href{https://doi.org/10.1007/JHEP02(2021)132}{\emph{JHEP} {\bfseries 02} (2021) 132} [\href{https://arxiv.org/abs/2008.05960}{{\ttfamily 2008.05960}}].

\bibitem{Apruzzi:2021nmk}
F.~Apruzzi, F.~Bonetti, I.n.~Garc\'\i{}a~Etxebarria, S.S.~Hosseini and S.~Schafer-Nameki, \emph{{Symmetry TFTs from String Theory}}, \href{https://doi.org/10.1007/s00220-023-04737-2}{\emph{Commun. Math. Phys.} {\bfseries 402} (2023) 895} [\href{https://arxiv.org/abs/2112.02092}{{\ttfamily 2112.02092}}].

\bibitem{Kong:2020cie}
L.~Kong, T.~Lan, X.-G.~Wen, Z.-H.~Zhang and H.~Zheng, \emph{{Algebraic higher symmetry and categorical symmetry -- a holographic and entanglement view of symmetry}}, \href{https://doi.org/10.1103/PhysRevResearch.2.043086}{\emph{Phys. Rev. Res.} {\bfseries 2} (2020) 043086} [\href{https://arxiv.org/abs/2005.14178}{{\ttfamily 2005.14178}}].

\bibitem{Freed:2022qnc}
D.S.~Freed, G.W.~Moore and C.~Teleman, \emph{{Topological symmetry in quantum field theory}},  \href{https://arxiv.org/abs/2209.07471}{{\ttfamily 2209.07471}}.

\bibitem{Freed:2022iao}
D.S.~Freed, \emph{{Introduction to topological symmetry in QFT.}}, \href{https://doi.org/10.1090/pspum/107/01946}{\emph{Proc. Symp. Pure Math.} {\bfseries 107} (2024) 93} [\href{https://arxiv.org/abs/2212.00195}{{\ttfamily 2212.00195}}].

\bibitem{Kaidi:2022cpf}
J.~Kaidi, K.~Ohmori and Y.~Zheng, \emph{{Symmetry TFTs for Non-invertible Defects}}, \href{https://doi.org/10.1007/s00220-023-04859-7}{\emph{Commun. Math. Phys.} {\bfseries 404} (2023) 1021} [\href{https://arxiv.org/abs/2209.11062}{{\ttfamily 2209.11062}}].

\bibitem{Burbano:2021loy}
I.M.~Burbano, J.~Kulp and J.~Neuser, \emph{{Duality defects in E$_{8}$}}, \href{https://doi.org/10.1007/JHEP10(2022)187}{\emph{JHEP} {\bfseries 10} (2022) 186} [\href{https://arxiv.org/abs/2112.14323}{{\ttfamily 2112.14323}}].

\bibitem{vanBeest:2022fss}
M.~van Beest, D.S.W.~Gould, S.~Schafer-Nameki and Y.-N.~Wang, \emph{{Symmetry TFTs for 3d QFTs from M-theory}}, \href{https://doi.org/10.1007/JHEP02(2023)226}{\emph{JHEP} {\bfseries 02} (2023) 226} [\href{https://arxiv.org/abs/2210.03703}{{\ttfamily 2210.03703}}].

\bibitem{Chen:2023qnv}
J.~Chen, W.~Cui, B.~Haghighat and Y.-N.~Wang, \emph{{SymTFTs and duality defects from 6d SCFTs on 4-manifolds}}, \href{https://doi.org/10.1007/JHEP11(2023)208}{\emph{JHEP} {\bfseries 11} (2023) 208} [\href{https://arxiv.org/abs/2305.09734}{{\ttfamily 2305.09734}}].

\bibitem{Sun:2023xxv}
Z.~Sun and Y.~Zheng, \emph{{When are Duality Defects Group-Theoretical?}},  \href{https://arxiv.org/abs/2307.14428}{{\ttfamily 2307.14428}}.

\bibitem{Apruzzi:2023uma}
F.~Apruzzi, F.~Bonetti, D.S.W.~Gould and S.~Schafer-Nameki, \emph{{Aspects of categorical symmetries from branes: SymTFTs and generalized charges}}, \href{https://doi.org/10.21468/SciPostPhys.17.1.025}{\emph{SciPost Phys.} {\bfseries 17} (2024) 025} [\href{https://arxiv.org/abs/2306.16405}{{\ttfamily 2306.16405}}].

\bibitem{Cao:2023rrb}
W.~Cao and Q.~Jia, \emph{{Symmetry TFT for subsystem symmetry}}, \href{https://doi.org/10.1007/JHEP05(2024)225}{\emph{JHEP} {\bfseries 05} (2024) 225} [\href{https://arxiv.org/abs/2310.01474}{{\ttfamily 2310.01474}}].

\bibitem{Duan:2023ykn}
Z.~Duan, Q.~Jia and S.~Lee, \emph{{\ensuremath{\mathbb{Z}}$_{N}$ duality and parafermions revisited}}, \href{https://doi.org/10.1007/JHEP11(2023)206}{\emph{JHEP} {\bfseries 11} (2023) 206} [\href{https://arxiv.org/abs/2309.01913}{{\ttfamily 2309.01913}}].

\bibitem{Antinucci:2024zjp}
A.~Antinucci and F.~Benini, \emph{{Anomalies and gauging of U(1) symmetries}}, \href{https://doi.org/10.1103/PhysRevB.111.024110}{\emph{Phys. Rev. B} {\bfseries 111} (2025) 024110} [\href{https://arxiv.org/abs/2401.10165}{{\ttfamily 2401.10165}}].

\bibitem{Brennan:2024fgj}
T.D.~Brennan and Z.~Sun, \emph{{A SymTFT for continuous symmetries}}, \href{https://doi.org/10.1007/JHEP12(2024)100}{\emph{JHEP} {\bfseries 12} (2024) 100} [\href{https://arxiv.org/abs/2401.06128}{{\ttfamily 2401.06128}}].

\bibitem{Bhardwaj:2024ydc}
L.~Bhardwaj, K.~Inamura and A.~Tiwari, \emph{{Fermionic Non-Invertible Symmetries in (1+1)d: Gapped and Gapless Phases, Transitions, and Symmetry TFTs}},  \href{https://arxiv.org/abs/2405.09754}{{\ttfamily 2405.09754}}.

\bibitem{Lan:2014uaa}
T.~Lan, J.C.~Wang and X.-G.~Wen, \emph{{Gapped Domain Walls, Gapped Boundaries and Topological Degeneracy}}, \href{https://doi.org/10.1103/PhysRevLett.114.076402}{\emph{Phys. Rev. Lett.} {\bfseries 114} (2015) 076402} [\href{https://arxiv.org/abs/1408.6514}{{\ttfamily 1408.6514}}].

\bibitem{Kaidi:2021gbs}
J.~Kaidi, Z.~Komargodski, K.~Ohmori, S.~Seifnashri and S.-H.~Shao, \emph{{Higher central charges and topological boundaries in 2+1-dimensional TQFTs}}, \href{https://doi.org/10.21468/SciPostPhys.13.3.067}{\emph{SciPost Phys.} {\bfseries 13} (2022) 067} [\href{https://arxiv.org/abs/2107.13091}{{\ttfamily 2107.13091}}].

\bibitem{Kobayashi:2022vgz}
R.~Kobayashi, \emph{{Symmetry-preserving boundary of (2+1)D fractional quantum Hall states}}, \href{https://doi.org/10.1103/PhysRevResearch.4.033137}{\emph{Phys. Rev. Res.} {\bfseries 4} (2022) 033137} [\href{https://arxiv.org/abs/2203.08156}{{\ttfamily 2203.08156}}].

\bibitem{Davydov:2010kfz}
A.~Davydov, M.~Mueger, D.~Nikshych and V.~Ostrik, \emph{{The Witt group of non-degenerate braided fusion categories}},  \href{https://arxiv.org/abs/1009.2117}{{\ttfamily 1009.2117}}.

\bibitem{Fuchs:2012dt}
J.~Fuchs, C.~Schweigert and A.~Valentino, \emph{{Bicategories for boundary conditions and for surface defects in 3-d TFT}}, \href{https://doi.org/10.1007/s00220-013-1723-0}{\emph{Commun. Math. Phys.} {\bfseries 321} (2013) 543} [\href{https://arxiv.org/abs/1203.4568}{{\ttfamily 1203.4568}}].

\bibitem{Davydov:2011dqx}
A.~Davydov, D.~Nikshych and V.~Ostrik, \emph{{On the structure of the Witt group of braided fusion categories}},  \href{https://arxiv.org/abs/1109.5558}{{\ttfamily 1109.5558}}.

\bibitem{Meir2011}
E.~Meir and E.~Musicantov, \emph{Module categories over graded fusion categories},  \href{https://arxiv.org/abs/1010.4333}{{\ttfamily 1010.4333}}.

\bibitem{Ostrik:2002ohv}
V.~Ostrik, \emph{{Module categories over the Drinfeld double of a finite group}},  \href{https://arxiv.org/abs/math/0202130}{{\ttfamily math/0202130}}.

\bibitem{Tsui:2017ryj}
L.~Tsui, Y.-T.~Huang, H.-C.~Jiang and D.-H.~Lee, \emph{{The phase transitions between $Z_n \times Z_n$ bosonic topological phases in 1 $+$ 1D, and a constraint on the central charge for the critical points between bosonic symmetry protected topological phases}}, \href{https://doi.org/10.1016/j.nuclphysb.2017.03.021}{\emph{Nucl. Phys. B} {\bfseries 919} (2017) 470} [\href{https://arxiv.org/abs/1701.00834}{{\ttfamily 1701.00834}}].

\bibitem{Seifnashri:2025fgd}
S.~Seifnashri, S.-H.~Shao and X.~Yang, \emph{{Gauging non-invertible symmetries on the lattice}},  \href{https://arxiv.org/abs/2503.02925}{{\ttfamily 2503.02925}}.

\bibitem{Warman:2024lir}
A.~Warman, F.~Yang, A.~Tiwari, H.~Pichler and S.~Schafer-Nameki, \emph{{Categorical Symmetries in Spin Models with Atom Arrays}},  \href{https://arxiv.org/abs/2412.15024}{{\ttfamily 2412.15024}}.

\bibitem{Seiberg:2024gek}
N.~Seiberg, S.~Seifnashri and S.-H.~Shao, \emph{{Non-invertible symmetries and LSM-type constraints on a tensor product Hilbert space}}, \href{https://doi.org/10.21468/SciPostPhys.16.6.154}{\emph{SciPost Phys.} {\bfseries 16} (2024) 154} [\href{https://arxiv.org/abs/2401.12281}{{\ttfamily 2401.12281}}].

\bibitem{Zhang:2024nbt}
H.-C.~Zhang and G.~Sierra, \emph{{Kramers-Wannier self-duality and non-invertible translation symmetry in quantum chains: a wave-function perspective}},  \href{https://arxiv.org/abs/2410.06727}{{\ttfamily 2410.06727}}.

\bibitem{Moradi:2023dan}
H.~Moradi, O.M.~Aksoy, J.H.~Bardarson and A.~Tiwari, \emph{{Symmetry fractionalization, mixed-anomalies and dualities in quantum spin models with generalized symmetries}}, \href{https://doi.org/10.21468/SciPostPhys.18.3.097}{\emph{SciPost Phys.} {\bfseries 18} (2025) 097} [\href{https://arxiv.org/abs/2307.01266}{{\ttfamily 2307.01266}}].

\bibitem{Seifnashri:2023dpa}
S.~Seifnashri, \emph{{Lieb-Schultz-Mattis anomalies as obstructions to gauging (non-on-site) symmetries}}, \href{https://doi.org/10.21468/SciPostPhys.16.4.098}{\emph{SciPost Phys.} {\bfseries 16} (2024) 098} [\href{https://arxiv.org/abs/2308.05151}{{\ttfamily 2308.05151}}].

\bibitem{Li:2023ani}
L.~Li, M.~Oshikawa and Y.~Zheng, \emph{{Noninvertible duality transformation between symmetry-protected topological and spontaneous symmetry breaking phases}}, \href{https://doi.org/10.1103/PhysRevB.108.214429}{\emph{Phys. Rev. B} {\bfseries 108} (2023) 214429} [\href{https://arxiv.org/abs/2301.07899}{{\ttfamily 2301.07899}}].

\bibitem{Kennedy:1992ifl}
T.~Kennedy and H.~Tasaki, \emph{{Hidden Z2\texttimes{}Z2 symmetry breaking in Haldane-gap antiferromagnets}}, \href{https://doi.org/10.1103/PhysRevB.45.304}{\emph{Phys. Rev. B} {\bfseries 45} (1992) 304}.

\bibitem{Kennedy:1992tke}
T.~Kennedy and H.~Tasaki, \emph{{Hidden symmetry breaking and the Haldane phase inS=1 quantum spin chains}}, \href{https://doi.org/10.1007/bf02097239}{\emph{Commun. Math. Phys.} {\bfseries 147} (1992) 431}.

\bibitem{M.Oshikawa_1992}
M.~Oshikawa, \emph{Hidden z2*z2 symmetry in quantum spin chains with arbitrary integer spin}, \href{https://doi.org/10.1088/0953-8984/4/36/019}{\emph{Journal of Physics: Condensed Matter} {\bfseries 4} (1992) 7469}.

\bibitem{Else:2014vma}
D.V.~Else and C.~Nayak, \emph{{Classifying symmetry-protected topological phases through the anomalous action of the symmetry on the edge}}, \href{https://doi.org/10.1103/PhysRevB.90.235137}{\emph{Phys. Rev. B} {\bfseries 90} (2014) 235137} [\href{https://arxiv.org/abs/1409.5436}{{\ttfamily 1409.5436}}].

\bibitem{Kawagoe:2021gqi}
K.~Kawagoe and M.~Levin, \emph{{Anomalies in bosonic symmetry-protected topological edge theories: Connection to F symbols and a method of calculation}}, \href{https://doi.org/10.1103/PhysRevB.104.115156}{\emph{Phys. Rev. B} {\bfseries 104} (2021) 115156} [\href{https://arxiv.org/abs/2105.02909}{{\ttfamily 2105.02909}}].

\bibitem{Sonia2017}
S.~Natale, \emph{{On the Equivalence of Module Categories over a Group-Theoretical Fusion Category}}, \href{https://doi.org/https://doi.org/10.3842/SIGMA.2017.042}{\emph{SIGMA} {\bfseries 13} (2017) 042} [\href{https://arxiv.org/abs/1608.04435}{{\ttfamily 1608.04435}}].

\end{thebibliography}\endgroup
\bibliographystyle{JHEP} 

\end{document}